\documentclass[11pt,a4paper]{article}
\usepackage{jheppub}

\usepackage{graphicx}
\usepackage[utf8]{inputenc}
\usepackage[english]{babel}

\usepackage{changepage}
\usepackage{color}
\usepackage{enumitem}
\usepackage{booktabs}
\usepackage{tabularx}
\usepackage{multirow}
\usepackage{array}

\usepackage{amssymb,latexsym,amsthm,amsmath,cancel}     
\usepackage{hyperref}

\newcommand{\bea}{\begin{eqnarray}}
\newcommand{\eea}{\end{eqnarray}}
\newcommand{\be}{\begin{equation}}
\newcommand{\ee}{\end{equation}}
\newcommand{\td}{\text{d}}
\newcommand{\te}{\text{e}}

\newcommand{\hx}{\hat{X}}
\newcommand{\hp}{\hat{\psi}}
\newcommand{\he}{\hat{\text{e}}}

 \title{Torsionful Killing–Yano Forms under T-Duality: Transformation Conditions and Emergent Isometries}

\author[a]{\"{O}zg\"{u}r Kelek\c{c}i}
\author[b]{\"{U}mit Ertem}
\author[c]{\"{O}zg\"{u}r A\c{c}{\i}k}

\affiliation[a]{Department of Basic Sciences, Faculty of Engineering,\\
 University of Turkish Aeronautical Association, 06790 Etimesgut, Ankara, Turkey}
\affiliation[b]{Department of Software Engineering, Ankara University,\\
Faculty of Engineering, 06830 G\"olba\c{s}{\i}, Ankara, Turkey}
\affiliation[c]{Department of Physics, Ankara University,\\ Faculty of Sciences, 06100, Tando\u gan-Ankara,
Turkey}

\emailAdd{okelekci@thk.edu.tr}
\emailAdd{umitertem@ankara.edu.tr}
\emailAdd{ozacik@science.ankara.edu.tr}

\abstract{We investigate the transformation of Killing--Yano (KY) \(p\)-forms under Abelian
T-duality in the presence of a non-trivial three-form torsion. Using a torsionful,
metric-compatible connection and the Buscher rules of T-duality, we derive compact,
coordinate-free transformation laws applicable to KY forms of arbitrary degree under
explicit matching assumptions. In particular, we show that a Killing 1-form is preserved
by the direct duality map whenever its transverse components are independent of the
isometry direction, with the dual circle component determined by a scalar correction
equation. The framework is then applied to several examples, including the Hopf
T-dual of \(S^3\), Schwarzschild spacetime, and the Nappi--Witten plane wave with exact
NS--NS torsion, where we analyze the transformed Killing--Yano forms explicitly.
Finally, we give a generalized-geometric and double-field-theoretic interpretation of
emergent Killing 1-forms in the T-dual geometry, showing how they arise from generalized
Killing data in the original duality frame.}

\begin{document}
\maketitle
\flushbottom

\section{Introduction}

Killing symmetries are among the most basic tools for extracting physical and geometric
information from a spacetime. They encode continuous isometries, lead to conserved
quantities along geodesics, and organize the invariant structure of a background. There
are, however, important symmetries which are not generated by Killing vector fields.
These hidden symmetries are described by higher-rank Killing tensors and antisymmetric
Killing--Yano forms. Killing--Yano tensors were introduced by Yano \cite{Yano1952} as
natural antisymmetric analogues of Killing vector fields. Their importance in gravitational
physics became especially clear through the study of the Kerr geometry, where Carter's
separation constant \cite{Carter1968} and the Walker--Penrose construction
\cite{WalkerPenrose1970} revealed that separability and integrability can be governed by
geometric structures beyond the ordinary isometry algebra.

Since then, Killing and Killing--Yano tensors have played a central role in the analysis
of classical and quantum dynamics on curved spaces. Symmetric Killing tensors generate
conserved quantities for geodesic motion, while Killing--Yano forms can be regarded as
more fundamental objects whose ``squares'' produce Killing tensors. They also enter the
construction of symmetry operators for wave equations and the Dirac equation, and they
appear naturally in spinning-particle models and worldline supersymmetry
\cite{PenroseRindler1986,Gibbons:1993ap,BennCharlton1997,BennKress2004,Cariglia2014}.
In higher-dimensional gravity, hidden symmetries generated by Killing--Yano and
conformal Killing--Yano tensors provide the organizing principle behind the complete
integrability of geodesic motion and the separability of field equations in many rotating
black-hole backgrounds
\cite{Frolov:2017kze,KubiznakFrolov2007,FrolovKubiznak2008,HouriOotaYasui2007,Krtous:2006qy}.
These developments show that Killing--Yano forms are not merely auxiliary
differential-geometric objects, but rather encode a refined symmetry structure of direct
relevance to gravitational and string-theoretic backgrounds.

In ordinary pseudo-Riemannian geometry, a Killing--Yano \(p\)-form is defined using the
Levi-Civita connection. In string theory and supergravity, however, the natural geometry
is usually not determined by the metric alone. The NS--NS two-form field \(B\), with field
strength \(H=\td B\), enters the low-energy theory together with the metric and dilaton,
and it naturally defines a skew-symmetric torsion contribution
\cite{Strominger1986,Hull1986}. This has led to generalized versions of Killing--Yano
equations in which the Levi-Civita connection is replaced by a metric-compatible
connection with totally antisymmetric torsion
\cite{Papadopoulos:2011,DeJonghe:1996ue}. Such torsionful hidden symmetries occur
in string backgrounds, in geometries with \(G\)-structure, and in black-hole solutions
such as the Kerr--Sen spacetime, where the torsion is identified with the NS--NS
three-form flux and is essential for the separability structure
\cite{Houri:2010,Houri:2012pq}. Thus, in the presence of a \(B\)-field, the torsionful
Killing--Yano equation is a natural string-geometric counterpart of the standard
Killing--Yano equation.

Another fundamental feature of string theory is T-duality. Abelian T-duality relates
backgrounds with an isometry by mixing the metric and \(B\)-field components along the
dualized direction. In its standard form, this transformation is encoded in the Buscher
rules \cite{Buscher:1987sk,Buscher:1987qj}, and its broader role in string theory has
been reviewed extensively in refs. \cite{Giveon,Alvar}. At the level of the NS--NS sector,
T-duality does not act only on the metric; it transforms the pair \((g,B)\) in a coupled
way. Consequently, any equation involving a torsionful connection built from \(g\) and
\(H=\td B\) is expected to transform nontrivially. The question of how Killing--Yano forms
behave under T-duality is therefore subtle: the duality changes the metric, the \(B\)-field,
the torsionful connection, and the splitting between the transverse geometry and the
circle direction.

The fact that T-duality does not automatically preserve geometric symmetries in their original form is already familiar from the study of supersymmetry. In ref. \cite{Kelekci}, the transformation of type II supergravity
supersymmetry variations under Abelian and non-Abelian T-duality was analyzed in detail.
A central lesson of that work is that supersymmetry preservation is controlled by the
dependence of Killing spinors on the dualized directions. More precisely, the relevant
condition is expressed through the Kosmann spinorial Lie derivative along the isometry
directions. If the Killing spinors are invariant in this spinorial sense, supersymmetry is
preserved under T-duality; if not, supersymmetry may be partially or completely lost. This
provides an important analogy for the present problem. For Killing--Yano forms, one
likewise expects that dependence on the dualized direction should control whether a hidden
symmetry survives as an ordinary Killing--Yano symmetry of the dual background.

The transformation of Killing and Killing--Yano tensors under string dualities was studied
systematically by Chervonyi and Lunin \cite{ChervonyiLunin}. Their analysis addresses
precisely the question of how hidden symmetries behave when one applies string dualities
to backgrounds such as higher-dimensional black holes. They showed that Killing vectors
and Killing tensors can be transformed under appropriate conditions, but that
Killing--Yano tensors require more care: the standard Killing--Yano equation is generally
not preserved in its original form. Instead, they introduced a modified Killing--Yano
equation adapted to the duality transformation and to the presence of the NS--NS sector.
In this way, ref. \cite{ChervonyiLunin} provides a duality-covariant approach to hidden
symmetries, in which the equation itself is modified so that the transformed object
satisfies the appropriate duality-compatible condition. This is one of the main points of
contact with the present work.

Our approach takes a different route, with a clear geometric
advantage. We retain the standard metric-compatible torsionful
Killing--Yano equation in both duality frames and determine instead how
a Killing--Yano $p$-form must transform so that the dual object remains
an ordinary torsionful Killing--Yano form. Rather than absorbing the
duality effects into a modified equation, we compare the original and
T-dual systems directly after decomposing the form into its transverse
and circle components. This isolates the obstruction to preservation
in explicit matching assumptions and compatibility equations for a
correction term in the transformed circle component. The result is
therefore not only a transformation law, but also a directly testable
survival criterion. For Killing 1-forms, this criterion becomes
particularly transparent: the transverse components must be independent
of the dualized direction, while the remaining component satisfies a
first-order scalar correction equation. This condition is the form
analogue of the duality-direction compatibility appearing in the
supersymmetry analysis of~\cite{Kelekci}.

There is also a natural generalized-geometric interpretation of these questions. In
generalized geometry, the tangent and cotangent bundles are combined into
\(TM\oplus T^*M\), and the metric and \(B\)-field are unified in an \(O(d,d)\)-covariant
structure \cite{Hitchin,Gualtieri}. Double field theory extends this viewpoint by
introducing doubled coordinates and a generalized metric on which T-duality acts linearly
\cite{Tseytlin1990,Tseytlin1991,Siegel1993,HullZwiebach,HohmHullZwiebach}. This framework is
particularly useful for understanding why ordinary symmetries may disappear or appear
after T-duality. A Killing vector which is not realized geometrically in one duality frame
may be encoded as part of a generalized vector, involving either a one-form gauge
parameter or dependence on a dual coordinate. After an \(O(d,d)\) transformation, the same
generalized object may become an ordinary Killing vector in the dual frame. This mechanism
is closely related to the double-field-theory interpretation of T-duality and of the
generalized Lie derivative \cite{Hohm:2013bwa,Aldazabal:2013sca,Berman:2013eva}.

The present paper combines these two perspectives. First, we derive direct transformation
conditions for torsionful Killing--Yano \(p\)-forms under Abelian T-duality. This gives a
survival criterion: it determines when an ordinary Killing--Yano form of the original
geometry is transformed into an ordinary Killing--Yano form of the T-dual geometry. Second,
for Killing 1-forms, we use generalized geometry and double field theory to discuss the
inverse problem: the origin of Killing vectors that are present in the dual geometry but
are not obtained from ordinary Killing vectors by the direct map. This second point is
important in examples such as the Hopf T-dual of \(S^3\), where the ordinary isometry
structure changes under T-duality.

The paper is organized as follows. In section~\ref{sec2}, we review Abelian T-duality in a coframe
adapted to the isometry direction. We write the metric and \(B\)-field in a form suitable
for applying the Buscher rules and derive the corresponding transformation of the
metric-compatible torsionful connection. The torsion used in the auxiliary connection is
allowed to be a general skew-symmetric three-form of the form \(H+\Psi\), where
\(H=\td B\) is the NS--NS flux and \(\Psi\) parametrizes additional skew-torsion freedom
used in the matching construction.

In section~\ref{sec3}, we apply these connection formulae to the torsionful Killing--Yano equation.
A general \(p\)-form is decomposed as \(K=\alpha+e^\theta\wedge\beta\), where \(\alpha\)
and \(\beta\) are transverse forms. We then compare the original and T-dual
Killing--Yano equations under explicit matching assumptions and derive the affine
transformation rule for the circle component. The correction term \(F\) is constrained by
compatibility equations, and the independent conditions are identified. The case \(p=1\)
is treated separately, leading to the transformation rule for Killing 1-forms and to the
scalar equation controlling the correction of the dual circle component. We also explain how retaining the standard torsionful Killing--Yano
equation distinguishes this construction from the duality-covariant
formulation of Chervonyi and Lunin~\cite{ChervonyiLunin}.

Section~\ref{sec4} contains examples. We first study the Hopf T-dual of \(S^3\). In this case, only
those Killing 1-forms whose transverse components are independent of the Hopf fibre
coordinate are transformed by the direct map, while the remaining original Killing 1-forms
fail the preservation condition. We also examine the Killing--Yano 2-forms on \(S^3\) and
show that the T-dual \(S^2\times S^1\) background admits no non-trivial torsionful
Killing--Yano 2-form for the torsionful connection determined by the dual NS--NS flux. We
then consider local Buscher transformations of the Schwarzschild solution along the
timelike and axial Killing directions. Finally, we analyze the Nappi--Witten plane wave
\cite{NappiWitten}, an exact WZW background with NS--NS torsion, where the T-dual
background remains locally of Nappi--Witten type and the torsionful Killing--Yano 2-forms
transform explicitly.

In section~\ref{sec5}, we turn to the generalized-geometric origin of Killing 1-forms that appear
in the dual geometry but are not captured by the direct Killing--Yano transformation of
section~\ref{sec3}. Using the generalized metric and the DFT generalized Lie derivative, we show
how such ordinary dual Killing vectors may arise from generalized Killing vectors in the
original doubled description. This construction is illustrated for the \(S^2\times S^1\)
dual of \(S^3\) and for the axial dual of the Schwarzschild background. Section~\ref{sec6}
summarizes the results and comments on possible extensions, including the relation to more
general duality-covariant treatments of hidden symmetries.

\section{Preliminaries on Abelian T-duality} 
\label{sec2}
We consider a spacetime with an Abelian isometry and choose coordinates
$\{x^{\mu'}\}=\{x^\mu,\theta\}$ adapted to the Killing direction, so that
$\mathcal{L}_{\partial_\theta}g=0$~\cite{Giveon}. The metric can then be
written in block form as
\begin{align}
g_{\mu' \nu'}  =  \left(
\begin{array}{cc}
 \ \bar{g}_{\mu \nu}+e^{2\sigma} A_\mu A_\nu \quad & e^{2\sigma} A_\mu \\  \
 e^{2\sigma} A_\nu \quad  & e^{2\sigma}  \\
\end{array}
\right)
\end{align}
where $\mu'=0,1,\ldots,n$, $\mu=0,1,\ldots,n-1$, and $\theta$
is the $n$th coordinate. The corresponding line element, $B$-field,
and adapted co-frame are
\begin{align} \label{beforeT}
    &\td s^2=e^{2 \sigma} \left(\text{d$\theta $}+A_\mu \td  x^{\mu}  \right)^2+\bar{g}_{\mu \nu} \td  x^{\mu}  \td  x^{\nu}  \ \\ &B=\overline{B}+b\wedge \text{d$\theta $}=\frac{1}{2}\overline{B}_{\mu \nu} \td  x^{\mu}\wedge  \td  x^{\nu} + B_\mu  \td  x^{\mu} \wedge 
 \text{d$\theta $}  \nonumber  \\
    &\te^{a}= \bar{\te}^{a}={\te_\mu}^a \td  x^{\mu}   \ , \quad \te^\theta=e^\sigma \left(\text{d$\theta $}+A_\mu \td  x^{\mu}   \right) \ , \quad a: 0,1,2,..., (n-1)\nonumber
\end{align}
\noindent Because $\theta$ generates an isometry, $A_\mu$, $\sigma$,
and $\bar g_{\mu\nu}$ are independent of $\theta$. We also assume
$\mathcal{L}_{\partial_\theta}B=0$, so that
$\overline B_{\mu\nu}$ and $B_\mu$ are $\theta$-independent.
Abelian T-duality is implemented by the Buscher rules
\cite{Buscher:1987sk,Buscher:1987qj}, which relate the original fields
$\{g_{\mu'\nu'},B_{\mu'\nu'}\}$ to the T-dual fields
$\{\widehat g_{\mu'\nu'},\widehat B_{\mu'\nu'}\}$.
\footnote{Hatted quantities denote T-dual quantities unless otherwise stated.}
\begin{align}\label{Buscher}
  \widehat{g}_{\mu \nu}&= g_{\mu \nu} - \frac{1}{g_{\theta \theta}} (g_{\mu \theta}g_{\nu \theta}-B_{\mu \theta}B_{\nu \theta})\ ,  \quad   \widehat{g}_{\mu \theta}= \frac{B_{\mu \theta}}{g_{\theta \theta}}  \ , \quad  \widehat{g}_{\theta \theta}=\frac{1}{g_{\theta \theta}}   \\
   \widehat{B}_{\mu \nu}&= B_{\mu \nu} - \frac{1}{g_{\theta \theta}} (B_{\mu \theta}g_{\nu \theta}-g_{\mu \theta}B_{\nu \theta}) \ ,  \quad   \widehat{B}_{\mu \theta}= \frac{g_{\mu \theta}}{g_{\theta \theta}}
\end{align}

\noindent Applying the Buscher rules to the metric and $B$-field gives
\begin{align}\label{afterT} 
    \widehat{\td s^2}&=e^{-2 \sigma} \left(\text{d$\theta $}+b \right)^2+\bar{g}_{\mu \nu} \text{d}x^{\mu} \text{d}x^{\nu}  \ , \quad \widehat{B}=\overline{B}+A\wedge (b+\text{d$\theta $})  \\
    \hat{\te}^{a}&= \bar{\te}^{a} \ , \quad \hat{\te}^\theta=e^{-\sigma} \left(\text{d$\theta $}+b \right) \nonumber
\end{align}

\noindent where $b=B_\mu \td x^\mu$ and $A=A_\mu \td x^\mu$. We compute the torsion 2-forms $T_{a'}$ and the corresponding
metric-compatible torsionful connection 1-forms $\omega_{a'b'}$.
In what follows, the label $(LC)$ is reserved only for Levi-Civita
quantities. In supergravity and string theory, the canonical NS--NS
torsion contribution is the 3-form field strength associated with the
2-form $B$-field, namely $H=\td B$~\cite{Strominger1986,Hull1986}. However, in the present work we allow the metric-compatible torsionful connection to have a more
general skew-torsion 3-form $\mathcal{T}=H+\Psi $ , where \(H\) is the NS--NS flux contribution and \(\Psi\) parametrizes
the additional skew-torsion freedom of the auxiliary connection. Thus $\Psi$ is
not introduced as an additional physical NS--NS flux, but as part of
the auxiliary torsionful connection used in the matching construction. With the convention
$T_{a'}=\frac12 T_{a'b'c'}e^{b'}\wedge e^{c'}$, one can write the
following relations.\footnote{\label{fn:index-convention}Latin indices are used for the non-coordinate frame basis, and primed indices $a'$ run over all frame directions.}
\begin{align}
H&=dB=d\overline{B}+e^{-\sigma}db\wedge e^\theta-A\wedge db, \label{eq:Hsplit}\\
\mathcal{T}&=H+\Psi,\qquad \psi_{a'}:=i_{X_{a'}}\Psi, \notag\\
T_a&=i_{X_a}\mathcal{T}=i_{X_a}H+\psi_a,\qquad
T_\theta=i_{X_\theta}\mathcal{T}=i_{X_\theta}H+\psi_\theta, \notag\\
\psi_a&=\frac12\psi_{akl}e^k\wedge e^l
       +\psi_{a\theta m}e^\theta\wedge e^m,
\qquad
\psi_\theta=\frac12\psi_{\theta kl}e^k\wedge e^l, \label{eq:Tpsi}\\
T_a&=i_{X_a}(d\overline{B}-A\wedge db)
     +e^{-\sigma}(i_{X_a}db)\wedge e^\theta+\psi_a, \label{eq:Ta}\\
T_\theta&=e^{-\sigma}db+\psi_\theta . \label{eq:Ttheta}
\end{align}

\noindent The antisymmetry condition on the components of $\psi_{a'}$ ensures that the forms $\psi_{a'}$ arise from the
globally defined skew 3-form $\Psi$\footnote{Equivalently, one may write
$\Psi=\frac1{3!}\psi_{a'b'c'}e^{a'}\wedge e^{b'}\wedge e^{c'}$,
so that $\psi_{a'}=i_{X_{a'}}\Psi$. The relation
$i_{X_{a'}}\psi_{b'}=-i_{X_{b'}}\psi_{a'}$ gives
$\psi_{a'b'c'}=-\psi_{b'a'c'}=-\psi_{a'c'b'}$; see \autoref{Appendix1}.}. Similarly, torsion 2-forms after T-duality can be computed  as 

\begin{align}
\widehat{H}
&=d\widehat{B}
=d\overline{B}-A\wedge db+e^\sigma dA\wedge \he ^\theta, \label{eq:Hhat}\\
\widehat{\mathcal{T}}
&=\widehat{H}+\widehat{\Psi},\qquad
\widehat{\psi}_{a'}:=i_{\widehat{X}_{a'}}\widehat{\Psi}, \notag\\
\widehat{T}_a
&=i_{\widehat{X}_a}\widehat{\mathcal{T}}
=i_{\widehat{X}_a}\widehat{H}+\widehat{\psi}_a
=i_{X_a}(d\overline{B}-A\wedge db)
+e^\sigma(i_{X_a}dA)\wedge \he^\theta+\widehat{\psi}_a, \label{eq:That_a}\\
\widehat{T}_\theta
&=i_{\widehat{X}_\theta}\widehat{\mathcal{T}}
=e^\sigma dA+\widehat{\psi}_\theta . \label{eq:That_theta}
\end{align}

\noindent For later use, we define
\begin{align}
h&:=\td\overline{B}-A\wedge\td b,
\qquad
\Lambda:=e^\sigma\td A+e^{-\sigma}\td b, \label{hlambda_def}\\
f_{ab}&
:=
\frac12\,i_{X_a}i_{X_b}
\left(e^{-\sigma}\td b-e^\sigma\td A\right). \label{fab_def}
\end{align}

\noindent We also denote by $\tau_{ab}$ the transverse connection
contribution determined by the co-frame $\te^a$,
\begin{align}\label{tau_def}
2\tau_{ab}
=
i_{X_b}\td\te_a
-i_{X_a}\td\te_b
+\left(i_{X_a}i_{X_b}\td\te_c\right)\te^c .
\end{align}

\noindent The detailed derivation of the connection formulas is given in Appendix~\ref{Appendix2}. The resulting relations between the
original torsionful connection one-forms $\omega_{a'b'}$ and their
T-dual counterparts $\widehat{\omega}_{a'b'}$ are

\begin{align} \label{wab_duals}
\hat{\omega}_{ab}&=\omega_{ab}+ \frac{1}{2} i_{X_a} i_{X_b} \biggl( e^{-\sigma}\td b + e^{\sigma}\td A\biggr) (\hat{\te}^\theta- \te^\theta) + \frac{1}{2} (i_{\hx_a} \hp_b-i_{X_a} \psi_b ) \nonumber \\
   \hat{\omega}_{\theta a}&= \omega_{\theta a} -(\td \sigma)_a (\te^\theta +\he^\theta)+\frac{1}{2} i_{X_a}  \biggl(2 e^{-\sigma} \td b -2 e^{\sigma} \td A + \psi_\theta  - \hp_\theta    \biggr) 
\end{align}

\noindent By using these connection 1-forms \eqref{wab_duals} we obtain relations between torsionful connections before ($\nabla$) and after T-duality ($\hat{\nabla}$) 

\begin{align}
       &\hat{\nabla}_{\hx_{a}} \he_{b} =  \nabla_{X_{a}} \te_{b} + f_{ab}(\te^\theta+\he^\theta) + \frac{1}{2} (i_{\hx_b} \hp_a-i_{X_b} \psi_a )\\
    &\hat{\nabla}_{\hx_{\theta}} \he_{a} =\nabla_{X_{\theta}} \te_{a}-(\td \sigma)_a(\te^\theta+\he^\theta)+\frac{1}{2}i_{X_a}( \hp_\theta-\psi_\theta) \\
    & \hat{\nabla}_{\hx_{a}} \he_{\theta} =-\nabla_{X_{a}} \te_{\theta}-\frac{1}{2}i_{X_a}( \hp_\theta+\psi_\theta) \\
    &\hat{\nabla}_{\hx_{\theta}} \he_{\theta} = -\nabla_{X_{\theta}} \te_{\theta} 
\end{align}

\noindent Finally, the original and T-dual orthonormal frame vectors are related by 
\begin{align}\label{basisvecs}
\hat{X}_a=X_a+e^{\sigma}  i_{X_a}(A-b)  X_\theta   \, ,   \quad   \hat{X}_\theta=e^{2\sigma} X_\theta
\end{align}

\section{T-duality on Killing--Yano equations} 
\label{sec3}
With the torsionful T-duality relations established, we now apply them
to the Killing--Yano equations. A $p$-form $\omega$ is a Killing-Yano $p$-form if
\begin{align}\label{KYano1}
    \nabla_X \omega = \frac{1}{p + 1} i_X d\omega
\end{align}
for all vector fields $X$, where $\nabla$ is the Levi-Civita connection. 
For a metric connection $\nabla^{\mathcal T}$ with totally
antisymmetric torsion 3-form $\mathcal T$, the corresponding
torsionful Killing--Yano equation is
\begin{align}\label{KYano2}
\nabla^{\mathcal T}_X\omega
=
\frac{1}{p+1}i_X\td^{\mathcal T}\omega,
\qquad
\td^{\mathcal T}\omega
:=
e^{a'}\wedge
\nabla^{\mathcal T}_{X_{a'}}\omega .
\end{align}
In the sequel, we use this torsionful form, since the T-duality
relations are naturally expressed in terms of torsionful connections.
In the notation of Section~\ref{sec2}, the torsion is the full skew 3-form
$\mathcal T=H+\Psi$. Using the index convention stated in
footnote~\ref{fn:index-convention}, and suppressing the superscript
$\mathcal T$ for notational simplicity, the torsionful
Killing--Yano equation can be written as

\begin{align}
p\,\nabla_{X_{a'}}\omega
=
-\,e^{b'}\wedge i_{X_{a'}}\!\big(\nabla_{X_{b'}}\omega\big).
\end{align}

\noindent We now split the full frame into transverse directions and the isometry direction. 
Let $K=\alpha+e^\theta\wedge\beta$, where $ \alpha $ is a transverse $p$-form and $ \beta $ is a transverse $(p-1)$-form. 

\noindent The torsionful Killing--Yano equation separates into four
component equations: the transverse and $e^\theta$-mixed parts of the
$X_a$ and $X_\theta$ equations. Substituting the original torsionful
connection components gives the following system for $K$.
The intermediate expansion is given in Appendix~\ref{Appendix2}.

\begin{equation}\label{KYorga1}
\begin{aligned}
0={}&p\Big(
X_a(\alpha)
+\Big(i_{X_a}\tau_c{}^{m}+\frac{1}{2}\psi_a{}^{m}{}_{c}\Big)e^c\wedge i_{X_m}\alpha
-\frac{1}{2}\big(i_{X_a}i_{X^{m}}h\big)\wedge i_{X_m}\alpha
-\Big(f_{ak}e^k+\frac{1}{2}i_{X_a}\psi_\theta\Big)\wedge\beta
\Big)
\\
&\quad
+e^b\wedge\Big(
i_{X_a}X_b(\alpha)
+\Big(i_{X_b}\tau_a{}^{m}+\frac{1}{2}\psi_b{}^{m}{}_{a}\Big)i_{X_m}\alpha
-\Big(i_{X_b}\tau_c{}^{m}+\frac{1}{2}\psi_b{}^{m}{}_{c}\Big)e^c\wedge i_{X_a}i_{X_m}\alpha
\\
&\qquad\qquad
-\frac{1}{2}i_{X_a}\Big[\big(i_{X_b}i_{X^{m}}h\big)\wedge i_{X_m}\alpha\Big]
-\Big(f_{ba}+\frac{1}{2}i_{X_a}i_{X_b}\psi_\theta\Big)\beta
+\Big(f_{bk}e^k+\frac{1}{2}i_{X_b}\psi_\theta\Big)\wedge i_{X_a}\beta
\Big).
\end{aligned}
\end{equation}

\begin{equation}\label{KYorga2}
\begin{aligned}
0={}&p\Big(
X_a(\beta)
+\Big(i_{X_a}\tau_c{}^{m}+\frac{1}{2}\psi_a{}^{m}{}_{c}\Big)e^c\wedge i_{X_m}\beta
-\frac{1}{2}\big(i_{X_a}i_{X^{m}}h\big)\wedge i_{X_m}\beta
-\Big(f_a{}^{m}-\frac{1}{2}\psi_{\theta a}{}^{m}\Big)i_{X_m}\alpha
\Big)
\\
&\quad
+e^b\wedge\Big(
i_{X_a}X_b(\beta)
+\Big(i_{X_b}\tau_a{}^{m}+\frac{1}{2}\psi_b{}^{m}{}_{a}\Big)i_{X_m}\beta
-\Big(i_{X_b}\tau_c{}^{m}+\frac{1}{2}\psi_b{}^{m}{}_{c}\Big)e^c\wedge i_{X_a}i_{X_m}\beta
\\
&\qquad\qquad
-\frac{1}{2}i_{X_a}\Big[\big(i_{X_b}i_{X^{m}}h\big)\wedge i_{X_m}\beta\Big]
-\Big(f_b{}^{m}-\frac{1}{2}\psi_{\theta b}{}^{m}\Big)i_{X_a}i_{X_m}\alpha
\Big)
\\
&\quad
+i_{X_a}X_\theta(\alpha)
+\frac{1}{2}i_{X_a}\Big[i_{X^{m}}(\Lambda+\psi_\theta)\wedge i_{X_m}\alpha\Big]
-(i_{X_a}d\sigma)\beta
+d\sigma\wedge i_{X_a}\beta.
\end{aligned}
\end{equation}

\begin{equation}\label{KYorgth1}
\begin{aligned}
0={}&p\,X_\theta(\alpha)-p\,d\sigma\wedge\beta
+i_{X^{m}}\Big[
\frac{p+1}{2}e^\sigma dA
+\frac{p-1}{2}\big(\psi_\theta+e^{-\sigma}db\big)
\Big]\wedge i_{X_m}\alpha
\\
&\quad
+e^b\wedge X_b(\beta)
+\Big(
\tau_c{}^{m}\wedge e^c
-i_{X^{m}}h
-\frac{1}{2}\psi^{m}{}_{bc}\,e^b\wedge e^c
\Big)\wedge i_{X_m}\beta.
\end{aligned}
\end{equation}

\begin{equation}\label{KYorgth2}
0=(p+1)\Big[X_\theta(\beta)
+
(i_{X^{m}}d\sigma)\,i_{X_m}\alpha
+\frac{1}{2}i_{X^{m}}(\Lambda+\psi_\theta)\wedge i_{X_m}\beta
\Big].
\end{equation}

\noindent For the T-dual geometry, we similarly write $\hat{K}
=
\hat{\alpha}
+
\hat{\te}^{\theta}\wedge\hat{\beta}$. Repeating the component decomposition with the T-dual
torsionful connection and then using \eqref{basisvecs} gives the following four T-dual Killing--Yano equations. The intermediate hatted equations
before substituting the transformed frame vectors are given in
Appendix~\ref{Appendix2}.

\begin{adjustwidth}{-1.8cm}{-1.8cm}
\begin{equation} \label{3.20}
\begin{aligned}
0={}&p\Big(
X_a(\hat\alpha)
+e^\sigma i_{X_a}(A-b)\,X_\theta(\hat\alpha)
+\Big(i_{X_a}\tau_c{}^{m}+\frac{1}{2}\hat\psi_a{}^{m}{}_{c}\Big)e^c\wedge i_{X_m}\hat\alpha
-\frac{1}{2}\big(i_{X_a}i_{X^{m}}h\big)\wedge i_{X_m}\hat\alpha
+\Big(f_{ak}e^k-\frac{1}{2}i_{X_a}\hat\psi_\theta\Big)\wedge\hat\beta
\Big)
\\
&\quad
+e^b\wedge\Big(
i_{X_a}X_b(\hat\alpha)
+\Big(i_{X_b}\tau_a{}^{m}+\frac{1}{2}\hat\psi_b{}^{m}{}_{a}\Big)i_{X_m}\hat\alpha
-\Big(i_{X_b}\tau_c{}^{m}+\frac{1}{2}\hat\psi_b{}^{m}{}_{c}\Big)e^c\wedge i_{X_a}i_{X_m}\hat\alpha
\\
&\qquad\qquad
-\frac{1}{2}i_{X_a}\Big[\big(i_{X_b}i_{X^{m}}h\big)\wedge i_{X_m}\hat\alpha\Big]
+\Big(f_{ba}-\frac{1}{2}i_{X_a}i_{X_b}\hat\psi_\theta\Big)\hat\beta
-\Big(f_{bk}e^k-\frac{1}{2}i_{X_b}\hat\psi_\theta\Big)\wedge i_{X_a}\hat\beta
\Big)
\\
&\quad
+e^\sigma (A-b)\wedge i_{X_a}X_\theta(\hat\alpha).
\end{aligned}
\end{equation}

\begin{equation}\label{3.21}
\begin{aligned}
0={}&p\Big(
X_a(\hat\beta)
+e^\sigma i_{X_a}(A-b)\,X_\theta(\hat\beta)
+\Big(i_{X_a}\tau_c{}^{m}+\frac{1}{2}\hat\psi_a{}^{m}{}_{c}\Big)e^c\wedge i_{X_m}\hat\beta
-\frac{1}{2}\big(i_{X_a}i_{X^{m}}h\big)\wedge i_{X_m}\hat\beta
+\Big(f_a{}^{m}+\frac{1}{2}\hat\psi_{\theta a}{}^{m}\Big)i_{X_m}\hat\alpha
\Big)
\\
&\quad
+e^b\wedge\Big(
i_{X_a}X_b(\hat\beta)
+\Big(i_{X_b}\tau_a{}^{m}+\frac{1}{2}\hat\psi_b{}^{m}{}_{a}\Big)i_{X_m}\hat\beta
-\Big(i_{X_b}\tau_c{}^{m}+\frac{1}{2}\hat\psi_b{}^{m}{}_{c}\Big)e^c\wedge i_{X_a}i_{X_m}\hat\beta
\\
&\qquad\qquad
-\frac{1}{2}i_{X_a}\Big[\big(i_{X_b}i_{X^{m}}h\big)\wedge i_{X_m}\hat\beta\Big]
+\Big(f_b{}^{m}+\frac{1}{2}\hat\psi_{\theta b}{}^{m}\Big)i_{X_a}i_{X_m}\hat\alpha
\Big)
\\
&\quad
+e^\sigma (A-b)\wedge i_{X_a}X_\theta(\hat\beta)
+e^{2\sigma}i_{X_a}X_\theta(\hat\alpha)
+\frac{1}{2}i_{X_a}\Big[i_{X^{m}}(\Lambda+\hat\psi_\theta)\wedge i_{X_m}\hat\alpha\Big]
+(i_{X_a}d\sigma)\hat\beta -d\sigma\wedge i_{X_a}\hat\beta.
\end{aligned}
\end{equation}

\begin{equation}\label{3.22}
\begin{aligned}
0={}&p\,e^{2\sigma}X_\theta(\hat\alpha)+p\,d\sigma\wedge\hat\beta
+i_{X^{m}}\Big[
\frac{p+1}{2}e^{-\sigma}db
+\frac{p-1}{2}\big(\hat\psi_\theta+e^\sigma dA\big)
\Big]\wedge i_{X_m}\hat\alpha
\\
&\quad
+e^b\wedge X_b(\hat\beta)
+e^\sigma (A-b)\wedge X_\theta(\hat\beta)
+\Big(
\tau_c{}^{m}\wedge e^c
-i_{X^{m}}h
-\frac{1}{2}\hat\psi^{m}{}_{bc}\,e^b\wedge e^c
\Big)\wedge i_{X_m}\hat\beta.
\end{aligned}
\end{equation}

\begin{equation}\label{3.23}
0=(p+1)\Big[e^{2\sigma}X_\theta(\hat\beta)
-(i_{X^{m}}d\sigma)\,i_{X_m}\hat\alpha
+\frac{1}{2}i_{X^{m}}(\Lambda+\hat\psi_\theta)\wedge i_{X_m}\hat\beta
\Big].
\end{equation}
\end{adjustwidth}

\noindent These two four-equation systems are the complete component
forms of the original and T-dual torsionful Killing--Yano equations.
They are not directly identical. The circle derivatives,
$\td\sigma$-dependent terms, and couplings to $\td A$, $\td b$,
$\psi_\theta$, and $\widehat{\psi}_\theta$ differ. Consequently, the
Buscher rules alone do not imply a direct preservation theorem for
Killing--Yano forms.

\noindent We now impose the following matching assumptions:

\begin{equation}\label{matching_conditions}
\begin{aligned}
&\quad \hat{\alpha}=\alpha,
\qquad
X_\theta(\alpha)=0,
\qquad
\psi_\theta=\hat{\psi}_\theta=-\Lambda
= -\left(e^\sigma \td A + e^{-\sigma}\td b\right),
\\[3pt]
&\left(i_{X_a}\tau_c{}^{m}+\frac12\,\psi_a{}^{m}{}_{c}\right)e^c
=
\frac12\,i_{X_a}i_{X^{m}}h,
\qquad \psi_a{}^{m}{}_{c}=\hat{\psi}_a{}^{m}{}_{c}
\end{aligned}
\end{equation}

\noindent These assumptions are not consequences of the Buscher rules.
They are sufficient matching conditions on the auxiliary
metric-compatible torsionful connections under which the original and T-dual component systems can be compared term by term. The comparison motivates the affine transformation ansatz

\begin{equation}\label{bsol}
\hat{\beta}=-e^{-2\sigma}\beta+F,
\qquad
X_\theta(F)=0,
\end{equation}

\noindent where $F$ is a transverse $(p-1)$-form. The homogeneous term
$-e^{-2\sigma}\beta$ is the expected inverse scaling of the circle
component under T-duality, while $F$ encodes the residual compatibility
conditions required for the hatted form to satisfy the standard
torsionful Killing--Yano equation.

\medskip

\noindent The detailed reduction of the matched systems is given in
Appendix~\ref{Appendix2}. The independent compatibility conditions
imposed on $F$ are the following. First,
\begin{equation}\label{F0cond}
\begin{aligned}
0={}&
(p-1)\Big(
e^\sigma i_{X_a}dA\wedge F
-
e^{-\sigma}i_{X_a}(dA+db)\wedge\beta
\Big)+2\Big(
e^{-\sigma}(dA+db)\wedge i_{X_a}\beta
-
e^\sigma dA\wedge i_{X_a}F
\Big).
\end{aligned}
\end{equation}
We denote this condition by $(F0)_a=0$. Second,

\begin{equation}\label{F2cond}
\begin{aligned}
0={}&
p\Big(
X_a(F)
+
e^{-\sigma}\Big(
i_{X_a}i_{X^m}(dA+db)
+
i_{X_a}(A-b)\,i_{X^m}d\sigma
\Big)i_{X_m}\alpha
\Big)
\\
&\quad
+e^b\wedge i_{X_a}X_b(F)
+e^{-\sigma}\Big(
i_{X^m}(dA+db)
+
(i_{X^m}d\sigma)(A-b)
\Big)\wedge i_{X_a}i_{X_m}\alpha
\\
&\quad
+(i_{X_a}d\sigma)F
-d\sigma\wedge i_{X_a}F
+2(p-1)e^{-2\sigma}(i_{X_a}d\sigma)\beta
+4e^{-2\sigma}d\sigma\wedge i_{X_a}\beta .
\end{aligned}
\end{equation}
We denote this condition by $(F2)_a=0$. There is also an intermediate relation, denoted $F1=0$, but it is not independent. Indeed, $e^a\wedge (F2)_a = F1$. Therefore, the independent conditions on the correction term $F$ are
\begin{equation}
(F0)_a=0,
\qquad
(F2)_a=0.
\end{equation}

\paragraph{Remark.}
There is also a special case in which the condition
$X_\theta(\alpha)=0$ can be relaxed. If $A=b$, while retaining the
remaining matching assumptions in \eqref{matching_conditions} but
relaxing $X_\theta(\alpha)=0$, and keeping
$\hat\beta=-e^{-2\sigma}\beta+F$ with $X_\theta(F)=0$, then the
independent conditions on $F$ are replaced by

\begin{align}
0={}&
(p-1)\left(
e^\sigma i_{X_a}dA\wedge F
-2e^{-\sigma}i_{X_a}dA\wedge\beta
\right)
+2\left(
2e^{-\sigma}dA\wedge i_{X_a}\beta
-e^\sigma dA\wedge i_{X_a}F
\right),
\notag\\[2mm]
0={}&
p\left(
X_a(F)
+2e^{-\sigma} i_{X_a}i_{X^m}dA\, i_{X_m}\alpha
\right)
+e^b\wedge i_{X_a}X_b(F)
+2e^{-\sigma} i_{X^m}dA\wedge i_{X_a}i_{X_m}\alpha
\notag\\
&+(i_{X_a}d\sigma)F
-d\sigma\wedge i_{X_a}F
+2(p-1)e^{-2\sigma}(i_{X_a}d\sigma)\beta
+4e^{-2\sigma}d\sigma\wedge i_{X_a}\beta
\notag\\
&+\left(e^{2\sigma}+e^{-2\sigma}\right)i_{X_a}X_\theta(\alpha).
\end{align}
These are the counterpart equations of \eqref{F0cond} and \eqref{F2cond} for \(A=b\) case. This case is very restrictive on the background, since it requires \(A=b\), but it
allows \(X_\theta(\alpha)\) to be nonzero.

\noindent Let us spell out the important special case of Killing 1-forms. For $p=1$, we write
\[
K=\alpha+e^\theta K_\theta,
\qquad
\alpha=K_a e^a,
\]
so that $\beta=K_\theta$ and the correction term $F$ is a function, which we denote by $f$. The transformation ansatz then reduces to
\begin{equation}\label{KoneTrans}
\hat K_a=K_a,
\qquad
X_\theta(K_a)=0,
\qquad
\hat K_\theta=-e^{-2\sigma}K_\theta+f,
\qquad
X_\theta(f)=0 .
\end{equation}
In this case, the condition \((F0)_a\) is identically satisfied, while \((F2)_a\) gives the remaining condition on $f$:
\begin{equation}\label{KoneFcond}
X_a(e^\sigma f)
=
K_m\Big(
i_{X^m}i_{X_a}(dA+db)
-
(i_{X^m}d\sigma)\,i_{X_a}(A-b)
\Big).
\end{equation}
Equation~\eqref{KoneFcond} determines $f$ locally up to the
homogeneous shift
\[
f\longmapsto f+C e^{-\sigma},
\]
where $C$ is constant. This freedom corresponds to adding a constant
multiple of the Killing 1-form associated with the dual circle.
Thus, under the matching conditions \eqref{matching_conditions}, a Killing 1-form is transformed to a T-dual Killing 1-form provided its transverse components are independent of the dualized isometry direction and the function $f$ satisfies \eqref{KoneFcond}. A similar compatibility with the dualized direction appears in the analysis of supersymmetry variations under T-duality, where Killing spinors were studied in detail~\cite{Kelekci}.

We now compare our construction more explicitly with the
duality-covariant approach of Chervonyi and Lunin~\cite{ChervonyiLunin}.
In their formulation, the transformation of a component with one leg
along the dualized circle contains the inverse scaling of the circle
factor. In our notation, this corresponds to the factor
$e^{-2\sigma}$ in the homogeneous part of \eqref{bsol}, up to the sign
convention associated with the chosen orthonormal frame. Their approach
achieves duality covariance by modifying the Killing--Yano equation and
imposing the corresponding background constraints.

The principal advantage of the present formulation is that the standard
torsionful Killing--Yano equation \eqref{KYano2} is retained unchanged
in both duality frames. Consequently, the transformed object remains an
ordinary torsionful Killing--Yano form, with its usual geometric
interpretation, rather than a solution of a modified equation. All
departures from the homogeneous circle scaling are isolated in the
correction term $F$ in
$\hat{\beta}=-e^{-2\sigma}\beta+F$. Equations
\eqref{F0cond} and \eqref{F2cond} therefore provide explicit and directly
testable criteria for the survival of the symmetry. In the regime
$\td A+\td b=0$, $\td\sigma=0$, and
$X_\theta(\alpha)=0$, one may take $F=0$, and the transformation reduces
to the pure homogeneous scaling.

\section{Examples} \label{sec4}
We now illustrate the above transformation conditions in some examples,
beginning with $S^3$ and the Schwarzschild solution. For the preserved
forms, we also display correction functions or forms compatible with
the transformation ansatz of Section~\ref{sec3}. All listed corrections
are independent of the dualized coordinate and satisfy
\eqref{KoneFcond} in the one-form case, and
\eqref{F0cond}--\eqref{F2cond} in the higher-degree case. The
homogeneous freedom is fixed so that the transformed forms agree with
the normalization used in the displayed dual bases. The resulting
hatted forms are also identified directly as Killing--Yano forms of
the corresponding dual backgrounds.

\subsubsection*{\underline{$S^3$ with no flux}} 
We employ Hopf coordinates $(\chi, \phi, \theta)$ to characterize $S^3$  with $\chi \in [0,\pi /2]$, $\phi \in [0,4 \pi]$, and  $\theta \in [0,2 \pi]$. The round metric on $S^3$ can be written as
\begin{align}\label{S3metric}
    \td s^2= \td\chi ^2+ \frac{1}{4}\left( \td \phi^2 +\td \theta^2-2\cos(2\chi) \td \phi \td \theta \right) \\
    \{\te^1,\te^2,\te^3\}=\{ \td\chi, \frac{1}{2}\sin(2\chi) \td \phi, \frac{1}{2}(\td \theta-\cos(2\chi) \td \phi) \}
\end{align}
We take the $B$-field to vanish in this example. The vector fields
$\partial_\phi$ and $\partial_\theta$ are Killing. We perform T-duality
along the $\theta$ direction and obtain the T-dual metric, $B$-field,
and torsion 3-form $H$:

\begin{align}\label{S3Tmetric}
    \widehat{\td s^2}&=\td\chi ^2+ \frac{1}{4}\sin^2(2\chi) \td \phi^2 + 4 \td \theta^2 \ ,\quad \{ \hat \te^1,\hat \te^2,\hat \te^3\}
=
\left\{
\td\chi,\,
\frac12\sin(2\chi)\td\phi,\,
2\td\theta
\right\}\nonumber \\
    \widehat{B}&= -\cos(2\chi) \td \phi \wedge \td \theta \ , \quad \widehat H=\td\widehat B=2\,\te^1 \wedge \te^2 \wedge \hat \te^3
\end{align}

\noindent The resulting T-dual metric \eqref{S3Tmetric} characterizes $S^2\times S^1$ product space, hence geometry and topology of the original space have changed.  Since $S^3$ is maximally symmetric, its six independent Killing vector fields give rise, by metric duality, to the following six Killing 1-forms:

\begin{align} \label{KillingS3}
K_{(i)}&=
\left(
\begin{array}{c}
\cos \theta\, \te^1 - \sin \theta\, \te^2 \\
\sin \theta\, \te^1 + \cos \theta\, \te^2 \\
\te^3 \\
\cos \phi\, \te^1 - \cos 2\chi \sin \phi\, \te^2 - \sin 2\chi \sin \phi\, \te^3 \\
\sin \phi\, \te^1 + \cos 2\chi \cos \phi\, \te^2 + \sin 2\chi \cos \phi\, \te^3 \\
\sin 2\chi\, \te^2 - \cos 2\chi\, \te^3
\end{array}
\right)& \\ \hat{K}_{(i)}&=\left(
\begin{array}{c}
\hat{\te}^3 \\
\cos \phi\, \te^1 - \cos 2\chi \sin \phi\, \te^2  \\
\sin \phi\, \te^1 + \cos 2\chi \cos \phi\, \te^2 \\
\sin 2\chi\, \te^2 
\end{array}
\right) \label{KillingS3Td}&
\end{align}

Killing 1-forms before and after T-duality are given in
\eqref{KillingS3} and \eqref{KillingS3Td}, respectively. The first two
original Killing 1-forms, $K_{(1)}$ and $K_{(2)}$, have transverse
components with explicit $\theta$-dependence and therefore fail the
condition $X_\theta(K_a)=0$ in \eqref{KoneTrans}. The remaining four 1-forms satisfy this condition. The correction functions reproducing the
dual basis in \eqref{KillingS3Td} are
\begin{align}
f_{(3)}&=5,
&
f_{(4)}&=-4\sin(2\chi)\sin\phi,
\nonumber\\
f_{(5)}&=4\sin(2\chi)\cos\phi,
&
f_{(6)}&=-4\cos(2\chi).
\end{align}
Direct substitution shows that these functions satisfy
\eqref{KoneFcond}, and \eqref{KoneTrans} gives
\begin{align}
K_{(3)}&\longmapsto\hat K_{(1)},
&
K_{(4)}&\longmapsto\hat K_{(2)},
\nonumber\\
K_{(5)}&\longmapsto\hat K_{(3)},
&
K_{(6)}&\longmapsto\hat K_{(4)} .
\end{align}

 We next consider KY 2-forms on the same background. The round $S^3$ admits four linearly independent KY 2-forms, which may be written as

\begin{align}\label{K2formsS3}
\Omega^{(1)}&=-\cos\chi\,\sin\psi\,\, e^{1}\wedge e^{2}
              -\sin\chi\,\cos\psi\,\, e^{2}\wedge e^{3}
              -\sin\chi\,\sin\psi\,\, e^{1}\wedge e^{3}, \nonumber \\ 
\Omega^{(2)}&=\phantom{-}\cos\chi\,\cos\psi\,\, e^{1}\wedge e^{2}
              -\sin\chi\,\sin\psi\,\, e^{2}\wedge e^{3}
              +\sin\chi\,\cos\psi\,\, e^{1}\wedge e^{3}, \nonumber \\ 
\Omega^{(3)}&=-\sin\chi\,\sin\zeta\,\, e^{1}\wedge e^{2}
              +\cos\chi\,\cos\zeta\,\, e^{2}\wedge e^{3}
              +\cos\chi\,\sin\zeta\,\, e^{1}\wedge e^{3}, \nonumber \\ 
\Omega^{(4)}&=\phantom{-}\sin\chi\,\cos\zeta\,\, e^{1}\wedge e^{2}
              +\cos\chi\,\sin\zeta\,\, e^{2}\wedge e^{3}
              -\cos\chi\,\cos\zeta\,\, e^{1}\wedge e^{3}.
\end{align}

\noindent where $\psi=\frac{\theta-\phi}{2}$ and $\zeta=\frac{\theta+\phi}{2}$. One immediately observes that $X_\theta(\alpha_i)\neq 0$ for the transverse 2-form components $\alpha_i$ of all KY 2-forms in \eqref{K2formsS3}. Hence, none of these KY 2-forms satisfies the preservation condition $X_\theta(\alpha)=0$, and therefore no dual KY 2-form is obtained from the original KY 2-forms listed in \eqref{K2formsS3} by the present T-duality map.

We now show that there does not exist any non-trivial torsionful KY 2-form on the dual geometry. Up to antisymmetry, the only nonzero Levi-Civita connection 1-form associated with the dual background \eqref{S3Tmetric} is
\[
(\hat\omega^{LC})^{1}{}_{2}=-2\cot(2\chi)\hat e^2,
\]
and the torsionful connection forms are
\[
\hat\omega^{1}{}_{2}=-2\cot(2\chi)\hat e^2+\hat e^3,
\qquad
\hat\omega^{1}{}_{3}=-\hat e^2,
\qquad
\hat\omega^{2}{}_{3}=\hat e^1
\]

\noindent with the remaining components determined by antisymmetry. In three dimensions, any 2-form can be written as \(\Omega=*v\), where  $v=v_a\hat e^a$. Since the torsionful connection is metric-compatible, the torsionful KY equation for \(\Omega\) is equivalent to
\begin{equation}
\nabla_{\hat X_a}v_b
=
\frac13\big(\nabla_{\hat X_c}v^c\big)\delta_{ab}.
\end{equation}
Thus the off-diagonal components of \(\nabla v\) vanish and the diagonal components are equal.

\noindent Using the above connection forms, the off-diagonal equations include
\[
\hat X_3(v_1)=-v_2,\qquad
\hat X_3(v_2)=v_1,
\qquad
\hat X_1(v_2)=-v_3,\qquad
\hat X_1(v_3)=v_2,
\]
while equality of the diagonal components gives
\[
\hat X_1(v_1)=\hat X_3(v_3).
\]
Since \(\hat X_3=\frac12\partial_\theta\), the first two equations imply
\[
v_1=A(\chi,\phi)\cos(2\theta)+B(\chi,\phi)\sin(2\theta),
\]
\[
v_2=A(\chi,\phi)\sin(2\theta)-B(\chi,\phi)\cos(2\theta).
\]
The equation \(\hat X_1(v_2)=-v_3\) then gives
\[
v_3=-A_\chi\sin(2\theta)+B_\chi\cos(2\theta).
\]
On the other hand,
\[
\hat X_1(v_1)=A_\chi\cos(2\theta)+B_\chi\sin(2\theta),
\]
whereas
\[
\hat X_3(v_3)=-A_\chi\cos(2\theta)-B_\chi\sin(2\theta).
\]
Therefore \(\hat X_1(v_1)=\hat X_3(v_3)\) implies
\[
A_\chi=B_\chi=0.
\]
Hence \(v_3=0\), and the equation \(\hat X_1(v_3)=v_2\) gives \(v_2=0\). Finally, \(\hat X_3(v_2)=v_1\) gives \(v_1=0\). Thus \(v=0\), and consequently \(\Omega=*v=0\). Therefore the T-dual background admits no non-trivial KY 2-form with respect to the torsionful connection determined by \((\hat g,\hat H)\).

\subsubsection*{\underline{Schwarzschild solution}}

Schwarzschild metric and the orthonormal co-frame can be written as
\begin{equation}
\begin{aligned}
ds^2
&=
-\left(1-\frac{2M}{r}\right)dt^2
+\left(1-\frac{2M}{r}\right)^{-1}dr^2
+r^2\left(d\theta^2+\sin^2\theta\,d\phi^2\right),
\\[2mm]
\{e^t,e^r,e^\phi,e^\theta\}
&=
\left\{
\sqrt{1-\frac{2M}{r}}\,dt,\,
\frac{dr}{\sqrt{1-\frac{2M}{r}}},\,
r\sin\theta\,d\phi,\,
r\,d\theta
\right\}
\end{aligned}
\end{equation}
The Schwarzschild spacetime has isometry group $
\mathbb{R}_t\times SO(3)$ corresponding to time translations and spatial rotations. Up to irrelevant overall signs,  a basis of Killing 1-forms is given by the metric duals of a basis of these four Killing vector fields:
\begin{equation}
K_{(i)}
=
\left(
\begin{array}{c}
\left(1-\frac{2M}{r}\right)^{1/2}\,e^t
\\[1mm]
r\sin\theta\,e^\phi
\\[1mm]
r\left(\cos\theta\cos\phi\,e^\phi+\sin\phi\,e^\theta\right)
\\[1mm]
r\left(\cos\phi\,e^\theta-\cos\theta\sin\phi\,e^\phi\right)
\end{array}
\right)
\end{equation}

\noindent Formally applying the same local Buscher transformation along the timelike
Killing direction \(t\), one obtains
\begin{equation}
d\hat{s}^2
=
-\left(1-\frac{2M}{r}\right)^{-1}dt^2
+\left(1-\frac{2M}{r}\right)^{-1}dr^2
+r^2\left(d\theta^2+\sin^2\theta\,d\phi^2\right),
\end{equation}
with
\[
\hat e^t
=
\left(1-\frac{2M}{r}\right)^{-1/2}dt,
\qquad
\hat e^r=e^r,
\qquad
\hat e^\phi=e^\phi,
\qquad
\hat e^\theta=e^\theta .
\]

All four Killing 1-forms have transverse components independent of the dualized coordinate \(t\). The transformed Killing 1-forms are

\begin{equation}
\hat K_{(i)}
=
\left(
\begin{array}{c}
\left(1-\frac{2M}{r}\right)^{-1/2}\hat e^t
\\[1mm]
r\sin\theta\, e^\phi
\\[1mm]
r\left(\cos\theta\cos\phi\, e^\phi+\sin\phi\,  e^\theta\right)
\\[1mm]
r\left(\cos\phi\,  e^\theta-\cos\theta\sin\phi\,  e^\phi\right)
\end{array}
\right)
\end{equation}

Writing \(\kappa(r)=1-\frac{2M}{r}\), the correction functions are
\begin{align}
f_{(1)}=\frac{2}{\sqrt{\kappa}},
\qquad
f_{(2)}=f_{(3)}=f_{(4)}=0 .
\end{align}
They satisfy \eqref{KoneFcond} and give exactly the normalization of
the four forms displayed above.

Thus T-duality along \(t\) preserves the full continuous isometry group \(\mathbb{R}_t\times SO(3)\). One can also apply T-duality along the \(\phi\)-direction, which results in the T-dual metric 
\begin{equation}
d\hat{s}^2
=
-\left(1-\frac{2M}{r}\right)dt^2
+\left(1-\frac{2M}{r}\right)^{-1}dr^2
+r^2d\theta^2
+\frac{1}{r^2\sin^2\theta}\,d\phi^2,
\end{equation}
with the transformed co-frame element in the $\phi$-direction
\[
\hat e^\phi=(r\sin\theta)^{-1}d\phi 
\]

Only the time-translation Killing 1-form and the axial rotational
Killing 1-form have transverse components satisfying
\(X_{\phi}(K_a)=0\). The transformed Killing 1-forms are
\begin{equation}
\hat K_{(i)}
=
\left(
\begin{array}{c}
\sqrt{1-\frac{2M}{r}}\,e^t
\\[1mm]
(r\sin\theta)^{-1}\hat e^\phi
\end{array}
\right)
\end{equation}

The corresponding correction functions are
\begin{align}
f_{(1)}=0,
\qquad
f_{(2)}=\frac{2}{r\sin\theta}.
\end{align}
Direct substitution verifies \eqref{KoneFcond}. These choices reproduce
the two dual forms with the normalization displayed above.
The remaining two rotational Killing 1-forms contain explicit \(\phi\)-dependent transverse components and are not transformed by the one-form T-duality map. Thus, on the local patch where the axial circle action is non-degenerate, T-duality along \(\phi\) preserves only the continuous isometry group $\mathbb{R}_t\times U(1)_\phi$.

\subsubsection*{\underline{Nappi--Witten plane wave with torsion}}

We next consider the four-dimensional Nappi--Witten plane wave, which is an exact WZW background based on a non-semisimple group \cite{NappiWitten}. In local coordinates $(u,w,\theta,x)$, we take
\begin{align}
    \td s^2
    =
    \td\theta^2+\td x^2-2\,\td u\,\td w
    +2x\,\td\theta\,\td u,
    \qquad
    B=u\,\td x\wedge\td\theta .
\end{align}
Hence the torsion 3-form is exact,
\begin{align}
    H=\td B=\td u\wedge\td x\wedge\td\theta .
\end{align}
The background is independent of $\theta$, and we apply Abelian T-duality along the $\theta$-direction. Rewriting the metric in the form adapted to this isometry gives
\begin{align}
    \td s^2
    =
    (\td\theta+x\,\td u)^2
    +\td x^2
    -2\,\td u\left(\td w+\frac{x^2}{2}\td u\right).
\end{align}
Thus, in the notation of Section~\ref{sec2},
\begin{align}
    \te^\theta=\td\theta+A,
    \qquad
    A=x\,\td u,
    \qquad
    b=u\,\td x,
    \qquad
    \sigma=0,
    \qquad
    \overline{B}=0.
\end{align}
Moreover,
\begin{align}
    \td A=\td x\wedge\td u,
    \qquad
    \td b=\td u\wedge\td x,
    \qquad
    \td A+\td b=0.
\end{align}
The Buscher rules then give the dual background
\begin{align}
    \td \hat{s}^2
    =
    (\td\theta+u\,\td x)^2
    +\td x^2
    -2\,\td u\left(\td w+\frac{x^2}{2}\td u\right),
    \qquad
    \hat B
    =
    x\,\td u\wedge(\td\theta+u\,\td x),
\end{align}
and therefore
\begin{align}
    \hat H=\td\hat B=-\td u\wedge\td x\wedge\td\theta .
\end{align}
This Abelian T-dual background is the standard Buscher dual of the Nappi--Witten plane wave \cite{MatlockParthasarathy}. It is locally again of Nappi--Witten form. Indeed, with
\begin{align}
    \Theta=\theta+ux,
    \qquad
    X=-x,
\end{align}
one obtains
\begin{align}
    \td\hat{s}^2
    =
    \td\Theta^2+\td X^2-2\,\td u\,\td w
    +2X\,\td\Theta\,\td u .
\end{align}

\noindent We now describe the torsionful Killing--Yano 2-forms. Introduce the following null co-frame, in which \(\ell^3\) and \(\ell^4\)
form a null pair
\begin{align}
    \ell^1&=\cos u\,\td\theta+\sin u\,\td x,
    &
    \ell^2&=-\sin u\,\td\theta+\cos u\,\td x,
    \nonumber\\
    \ell^3&=\td u,
    &
    \ell^4&=\td w-x\,\td\theta .
\end{align}
Then
\begin{align}
    \td s^2=(\ell^1)^2+(\ell^2)^2-2\ell^3\ell^4,
    \qquad
    H=-\ell^3\wedge\ell^1\wedge\ell^2 .
\end{align}
With this sign convention, \(H\) defines the parallelizing skew torsion
of the Nappi--Witten WZW background. Equivalently, the torsionful
metric-compatible connection \(\nabla^H\) used in the previous sections,
with torsion determined by the 3-form \(H\), satisfies
\begin{align}
    \nabla^H\ell^a=0,
    \qquad
    a=1,2,3,4.
\end{align}
It follows that every constant-coefficient 2-form constructed from the
\(\ell^a\) is torsionfully parallel and hence satisfies the torsionful
Killing--Yano equation. A convenient basis is
\begin{align}
    \Omega_1&=\ell^1\wedge\ell^2,
    &
    \Omega_2&=\ell^1\wedge\ell^3,
    &
    \Omega_3&=\ell^1\wedge\ell^4,
    \nonumber\\
    \Omega_4&=\ell^2\wedge\ell^3,
    &
    \Omega_5&=\ell^2\wedge\ell^4,
    &
    \Omega_6&=\ell^3\wedge\ell^4 .
\end{align}
Indeed, since \(\nabla^H\ell^a=0\), the components of each \(\Omega_i\)
in the \(\ell^a\) co-frame are constant and
\begin{align}
    \nabla^H_a(\Omega_i)_{bc}=0 .
\end{align}
Therefore
\begin{align}
    \nabla^H_a(\Omega_i)_{bc}
    +
    \nabla^H_b(\Omega_i)_{ac}
    =
    0,
    \qquad
    i=1,\ldots,6,
\end{align}
so each \(\Omega_i\) satisfies the torsionful Killing--Yano 2-form
equation.

We now apply the transformation conditions derived above. Since
\begin{align}
    e^\theta=\td\theta+x\,\td u,
    \qquad
    \hat e^\theta=\td\theta+u\,\td x,
\end{align}
each \(\Omega_i\) can be decomposed as
\begin{align}
    \Omega_i=\alpha_i+e^\theta\wedge\beta_i .
\end{align}
The transverse parts \(\alpha_i\) are independent of the dualized
coordinate,
\begin{align}
    X_\theta(\alpha_i)=0,
    \qquad
    i=1,\ldots,6.
\end{align}

Furthermore, since \(\sigma=0\) and \(\td A+\td b=0\), the correction
forms for all six representatives are
\begin{align}
F_{(i)}=0,
\qquad
i=1,\ldots,6.
\end{align}
These satisfy \eqref{F0cond} and \eqref{F2cond}. Hence
\begin{align}
\hat\beta_i=-\beta_i,
\qquad
\hat\Omega_i=\alpha_i-\hat e^\theta\wedge\beta_i,
\qquad
i=1,\ldots,6.
\end{align}

To verify explicitly that the transformed forms are torsionful
Killing--Yano forms of the dual geometry, introduce the dual null
co-frame
\begin{align}
    \hat\ell^1&=\cos u\,\td\Theta+\sin u\,\td X,
    &
    \hat\ell^2&=-\sin u\,\td\Theta+\cos u\,\td X,
    \nonumber\\
    \hat\ell^3&=\td u,
    &
    \hat\ell^4&=\td w-X\,\td\Theta .
\end{align}
Then
\begin{align}
    \td\hat s^2
    =
    (\hat\ell^1)^2+(\hat\ell^2)^2
    -2\hat\ell^3\hat\ell^4,
    \qquad
    \hat H=-\hat\ell^3\wedge\hat\ell^1\wedge\hat\ell^2 .
\end{align}
Thus the dual torsionful metric-compatible connection
\(\nabla^{\hat H}\), with torsion determined by \(\hat H\), is again
parallelizing,
\begin{align}
    \nabla^{\hat H}\hat\ell^a=0,
    \qquad
    a=1,2,3,4.
\end{align}
The transformed forms are
\begin{align}
    \hat\Omega_1&=\hat\ell^1\wedge\hat\ell^2,
    &
    \hat\Omega_2&=-\hat\ell^1\wedge\hat\ell^3,
    &
    \hat\Omega_3&=-\hat\ell^1\wedge\hat\ell^4,
    \nonumber\\
    \hat\Omega_4&=-\hat\ell^2\wedge\hat\ell^3,
    &
    \hat\Omega_5&=-\hat\ell^2\wedge\hat\ell^4,
    &
    \hat\Omega_6&=\hat\ell^3\wedge\hat\ell^4 .
\end{align}
Since each \(\hat\ell^a\) is torsionfully parallel, each
\(\hat\Omega_i\) has constant components in the dual null co-frame and
satisfies
\begin{align}
    \nabla^{\hat H}_a(\hat\Omega_i)_{bc}
    +
    \nabla^{\hat H}_b(\hat\Omega_i)_{ac}
    =
    0,
    \qquad
    i=1,\ldots,6.
\end{align}
Therefore all six torsionful Killing--Yano 2-forms above transform into
torsionful Killing--Yano 2-forms of the T-dual Nappi--Witten background.
The main outcomes of the examples discussed above are summarized in
Table~\ref{tab:example_summary}.

\begin{table}[t]
\centering
\small
\setlength{\tabcolsep}{5pt}
\renewcommand{\arraystretch}{1.18}
\begin{tabularx}{\textwidth}{
@{}
>{\raggedright\arraybackslash}p{3.2cm}
>{\centering\arraybackslash}p{1.8cm}
>{\raggedright\arraybackslash}p{3.3cm}
>{\raggedright\arraybackslash}X
@{}
}
\toprule
\textbf{Original background}
&
\textbf{T-dual direction}
&
\textbf{Original KY content}
&
\textbf{T-dual background and surviving KY forms}
\\
\midrule

$S^3$ with $B=0$
&
$\theta$
&
6 Killing 1-forms; 4 Killing--Yano 2-forms
&
The dual geometry is $S^2\times S^1$. \newline
$4/6$ Killing 1-forms survive,
while none of the four Killing--Yano 2-forms survives.
\\

\addlinespace[2pt]

\multirow{2}{3.2cm}{\raggedright Schwarzschild with $B=0$}
&
$t$
&
4 Killing 1-forms
&
All $4/4$ Killing 1-forms survive, and the full continuous
isometry algebra is preserved:
$\mathbb{R}_t\times SO(3)
\longrightarrow
\mathbb{R}_t\times SO(3)$.
\\

\cmidrule(l){2-4}

&
$\phi$
&
4 Killing 1-forms
&
Only $2/4$ Killing 1-forms survive, corresponding to time
translation and axial rotation:
$\mathbb{R}_t\times SO(3)
\longrightarrow
\mathbb{R}_t\times U(1)_\phi$.
\\

\addlinespace[2pt]

Nappi--Witten plane wave with $B\neq 0$
&
$\theta$ 
&
6 torsionful Killing--Yano 2-forms
&
The dual background is locally again of Nappi--Witten type,
and all six torsionful Killing--Yano 2-forms survive.
\\

\bottomrule
\end{tabularx}
\caption{Summary of the examples discussed in
section~\ref{sec4}. Which Killing--Yano forms survive depends on both
the background and the chosen T-duality direction.}
\label{tab:example_summary}
\end{table}

\section{Generalized--geometric origin of emergent Killing 1-forms} \label{sec5}

The transformation conditions derived in the previous sections determine when an ordinary Killing--Yano form of the original background transforms into a Killing--Yano form of the T-dual background. They therefore solve a survival problem. They do not, by themselves, solve the inverse problem of detecting Killing--Yano forms which appear only after T-duality.

For Killing 1-forms, this inverse problem can be formulated more sharply. A Killing 1-form is the metric dual of an ordinary Killing vector. Therefore, instead of working directly with 1-forms, one may work with the vector fields generating the corresponding isometries. Generalized geometry combines ordinary vector fields and one-form gauge parameters into sections of $TM\oplus T^*M$ \cite{Hitchin,Gualtieri}. Double field theory makes this structure $O(d,d)$-covariant by introducing doubled coordinates
\begin{align}
    X^M=(x^i,\tilde x_i),
\end{align}
together with the generalized metric $\mathcal H(g,B)$ and the DFT generalized Lie derivative \cite{HullZwiebach,HohmHullZwiebach}. For a background $(g,B)$, the generalized metric is
\begin{align}
    \mathcal H(g,B)
    =
    \begin{pmatrix}
        g-Bg^{-1}B & Bg^{-1}\\[1mm]
        -g^{-1}B & g^{-1}
    \end{pmatrix}.
\end{align}
A generalized vector is written as
\begin{align}
    \mathcal V=v+\lambda
    =
    v^i\partial_i+\lambda_i \td x^i,
\end{align}
or equivalently $\mathcal V^M=(v^i,\lambda_i)$. We call  $\mathcal V$ a generalized Killing vector \footnote{Since our aim here is only to reconstruct ordinary Killing vectors of the T-dual metric, we impose the generalized Killing condition on \(\mathcal H\) only. Including the generalized dilaton would give the corresponding symmetry condition for the full DFT background.} if 
\begin{align}
    \widehat{\mathcal L}_{\mathcal V}\mathcal H=0,
\end{align}
where
\begin{align}
    (\widehat{\mathcal L}_{\mathcal V}\mathcal H)_{MN}
    &=
    \mathcal V^P\partial_P\mathcal H_{MN}
    +
    \left(\partial_M\mathcal V^P-\partial^P \mathcal V_M\right)\mathcal H_{PN}
    +
    \left(\partial_N\mathcal V^P-\partial^P \mathcal V_N\right)\mathcal H_{MP}.
\end{align}
Indices are raised and lowered with the $O(d,d)$ metric. In a geometric section, where all fields and parameters depend only on the ordinary coordinates, this equation reduces to
\begin{align}
    \mathcal L_v g=0,
    \qquad
    \mathcal L_v B+\td\lambda=0.
\end{align}
Thus $v$ is an ordinary Killing vector of $g$, while $\lambda$ compensates the possible variation of $B$ by a gauge transformation.
The doubled description is subject to the strong constraint
\cite{HullZwiebach,ZwiebachLectures}. For any field or gauge parameter \(\Phi\), and for any pair of fields or gauge parameters $(\Phi,\Phi')$, one imposes
\begin{align}
    \eta^{MN}\partial_M\partial_N \Phi=0,
    \qquad
    \eta^{MN}\partial_M\Phi\,\partial_N\Phi'=0.
\end{align}

The usual geometric section corresponds to $\tilde\partial^i(\cdot)=0$. In the applications below, however, the background is independent of the dualized coordinate $y$. We therefore allow generalized Killing parameters to depend on the corresponding dual coordinate $\tilde y$, while keeping $\partial_y(\cdot)=0$. Since no object depends simultaneously on $y$ and $\tilde y$, and since no dependence on the other dual coordinates is introduced, the strong constraint remains satisfied in this restricted sector. This is the sector in which non-geometric generalized symmetries of the original background become ordinary geometric symmetries after T-duality. More explicitly, in this restricted sector the background satisfies
\(\tilde\partial^i\mathcal H=0\) and \(\partial_y\mathcal H=0\), while
the generalized Killing parameter may satisfy
\(\tilde\partial^y\mathcal V\neq0\), with
\(\partial_y\mathcal V=0\) and \(\tilde\partial^\mu\mathcal V=0\).
Hence all contractions appearing in the strong constraint vanish.

\paragraph{Proposition.}
Let $\mathcal V$ be a generalized Killing vector of the original generalized metric,
\begin{align}
    \widehat{\mathcal L}_{\mathcal V}\mathcal H=0.
\end{align}
Let $O_y\in O(d,d)$ denote the factorized T-duality transformation along the isometry direction $y$, and define
\begin{align}
    \widehat{\mathcal H}=O_y\mathcal H O_y^T,
    \qquad
    \widehat{\mathcal V}=O_y\mathcal V.
\end{align}
Then
\begin{align}
    \widehat{\mathcal L}_{\widehat{\mathcal V}}\widehat{\mathcal H}=0.
\end{align}
If $\widehat{\mathcal V}$ is geometric in the T-dual description, namely
\begin{align}
    \widehat{\mathcal V}=\hat v+\hat\lambda
\end{align}
with components depending only on the T-dual physical coordinates, then
\begin{align}
    \mathcal L_{\hat v}\hat g=0,
    \qquad
    \mathcal L_{\hat v}\hat B+\td\hat\lambda=0.
\end{align}
In particular, $\hat v$ is an ordinary Killing vector of the T-dual metric $\hat g$, and
\begin{align}
    \hat K=\hat g(\hat v,\cdot)
\end{align}
is a Killing 1-form on the T-dual background.

\paragraph{Proof.}
The generalized metric and the generalized Lie derivative transform covariantly under the $O(d,d)$ transformation associated with T-duality. Hence the generalized Killing equation is preserved by $O_y$. If $\widehat{\mathcal V}$ is geometric in the T-dual description, the generalized Killing equation decomposes into
\begin{align}
    \mathcal L_{\hat v}\hat g=0,
    \qquad
    \mathcal L_{\hat v}\hat B+\td\hat\lambda=0.
\end{align}
The first of these is the ordinary Killing equation for $\hat v$. Therefore its metric dual $\hat K=\hat g(\hat v,\cdot)$ is a Killing 1-form.

The purpose of the construction is to identify how ordinary Killing
vectors of the T-dual metric are encoded in the original generalized
metric before passing to the dual geometric frame. Generalized Killing
vectors which are non-geometric in the original description, due to
one-form components or dependence on the dual coordinate, may become
ordinary Killing vector fields after the \(O(1,1)\) exchange associated
with T-duality.

The proposition suggests a practical way to search for emergent dual Killing vectors. Let \(y\) be the T-dualized coordinate and let \(x^\mu\) denote the remaining coordinates, so that
\begin{align}
    x^i=(x^\mu,y).
\end{align}
A generalized vector decomposes as
\begin{align}
    \mathcal V
    =
    v^\mu\partial_\mu
    +
    v^y\partial_y
    +
    \lambda_\mu \td x^\mu
    +
    \lambda_y \td y .
\end{align}
Under T-duality along \(y\), the vector component along \(y\) and the
one-form component along \(\td y\) are exchanged,
\begin{align}
    v^y
    \longleftrightarrow
    \lambda_y .
\end{align}
The transverse vector components remain vector components. Therefore,
to obtain an ordinary vector field after T-duality, it is natural to
consider generalized vectors of the form
\begin{align}
    \mathcal V
    =
    v^\mu(x,\tilde y)\partial_\mu
    +
    q(x,\tilde y)\,\td y .
    \label{dual_geometric_ansatz}
\end{align}
Indeed, after T-duality this projects to
\begin{align}
    \hat v
    =
    v^\mu(x,\hat y)\partial_\mu
    +
    q(x,\hat y)\partial_{\hat y},
\end{align}
where the original dual coordinate \(\tilde y\) is reinterpreted as the
physical T-dual coordinate \(\hat y\). Thus the ansatz
\eqref{dual_geometric_ansatz} is dictated by the \(O(1,1)\) action:
the component \(q\,\td y\) is precisely the component which becomes
\(q\,\partial_{\hat y}\) after T-duality. This observation will be used below as the basis of our ansatz-selection method for generalized vectors.

The functions \(v^\mu\) and \(q\) are not specified a priori. They are
determined by the generalized Killing equation
\begin{align}
    \widehat{\mathcal L}_{\mathcal V}\mathcal H=0 .
\end{align}
In the examples below, the compatibility conditions force harmonic
dependence on the dual coordinate, giving the real solutions
\(\cos\tilde y\) and \(\sin\tilde y\). Hence the trigonometric
dependence is a consequence of the generalized Killing equation, not
an additional assumption.

Let \(\mathfrak G_y(\mathcal H)\) denote the space of solutions of the
generalized Killing equation within the ansatz
\begin{align}
    \mathcal V=v^\mu\partial_\mu+q\,\td y .
\end{align}
Define the T-dual projection map
\begin{align}
    \mathcal P_y:\mathfrak G_y(\mathcal H)\longrightarrow \Gamma(T\widehat M)
\end{align}
by
\begin{align}
    \mathcal P_y(\mathcal V)
    =
    v^\mu(x,\hat y)\partial_\mu
    +
    q(x,\hat y)\partial_{\hat y}.
\end{align}
By the proposition, \(\mathcal P_y(\mathcal V)\) is an ordinary Killing
vector of the T-dual metric whenever the projected object is geometric
in the T-dual description. Since the transformation law derived in
Section~\ref{sec3} is formulated for Killing 1-forms, we compare the resulting
vector fields with the surviving part by metric duality. Let
\(\mathfrak{isom}_{\rm surv}(\hat g)\subset \Gamma(T\widehat M)\)
denote the subspace of T-dual Killing vector fields whose metric-dual
1-forms belong to the image of the one-form transformation map of
Section~\ref{sec3}. Equivalently,
\begin{align}
    \hat v\in\mathfrak{isom}_{\rm surv}(\hat g)
    \quad\Longleftrightarrow\quad
    \hat v^\flat:=\hat g(\hat v,\cdot)
    \text{ is a surviving T-dual Killing 1-form.}
\end{align}
The emergent vector-field sector is then measured by
\begin{align}
    \frac{
    \operatorname{im}\mathcal P_y
    }{
    \operatorname{im}\mathcal P_y\cap
    \mathfrak{isom}_{\rm surv}(\hat g)
    }
\end{align}
Taking the metric duals \(\hat v^\flat=\hat g(\hat v,\cdot)\) of representatives of this quotient gives the corresponding emergent Killing 1-forms. Thus the dimension of this quotient gives the number of new Killing vectors, or equivalently new Killing 1-forms after metric duality, obtained by this generalized construction, provided the chosen solution space \(\mathfrak G_y(\mathcal H)\) has been fully solved in the desired function class.

\subsection{\texorpdfstring{$S^2\times S^1$ example}{S2 x S1 example}}

We now apply the construction to the reverse of the $S^3$ example considered earlier. Take the original background to be
\begin{align}
    \td s^2
    =
    \td\chi^2+\frac14\sin^2(2\chi)\td\phi^2+4\td\theta^2,
\end{align}
with
\begin{align}
    B=-\cos(2\chi)\,\td\phi\wedge\td\theta.
\end{align}
This is the T-dual of the round $S^3$ background. We now T-dualize along $y=\theta$. The T-dual metric is
\begin{align}
    \td\hat s^2
    =
    \td\chi^2
    +
    \frac14
    \left(
    \td\phi^2+\td\hat\theta^2
    -2\cos(2\chi)\td\phi\,\td\hat\theta
    \right),
\end{align}
which is the round $S^3$ metric. The ordinary original geometry has four Killing vectors, whereas the dual round $S^3$ has six. We now recover the two extra ones from the original generalized metric.
The following reconstruction is local on the coordinate patch where
\(\sin(2\chi)\neq 0\). The factors such as \(\csc(2\chi)\) and
\(\cot(2\chi)\) which appear below reflect the use of Hopf coordinates and the
degeneration loci of the angular parametrization. They do not affect the local
generalized-Killing interpretation of the dual Killing vectors.

For \(y=\theta\), the ansatz-selection method described above leads us to generalized vectors whose transverse vector components lie along the \((\chi,\phi)\)-directions and whose one-form component lies along \(\td\theta\). Thus we take
\begin{align}
    \mathcal V
    =
    F(\chi,\tilde\theta)\partial_\chi
    +
    G(\chi,\tilde\theta)\partial_\phi
    +
    Q(\chi,\tilde\theta)\td\theta.
\end{align}
This is precisely the inverse \(O(1,1)\) image of an ordinary vector
field in the T-dual description,
\begin{align}
    \hat v
    =
    \hat v^\chi\partial_\chi
    +
    \hat v^\phi\partial_\phi
    +
    \hat v^{\hat\theta}\partial_{\hat\theta}.
\end{align}

Substituting the ansatz into
\begin{align}
    \widehat{\mathcal L}_{\mathcal V}\mathcal H=0
\end{align}
gives the system
\begin{align}
    \partial_\chi F&=0, \\
    \partial_\chi Q
    &=
    -4\csc^2(2\chi)\,\partial_{\tilde\theta}F, \\
    \partial_\chi G
    &=
    \cos(2\chi)\,\partial_\chi Q, \\
    \partial_{\tilde\theta}Q
    &=
    -2F\cot(2\chi), \\
    \partial_{\tilde\theta}G
    &=
    -2F\csc(2\chi).
\end{align}
The compatibility conditions imply
\begin{align}
    \partial_{\tilde\theta}^2F+F=0.
\end{align}
Thus the two real independent nonzero modes are
\begin{align}
    F=\cos\tilde\theta,
    \qquad
    F=\sin\tilde\theta.
\end{align}
A convenient way to write the corresponding solutions is
\begin{align}
    G=2\,\partial_{\tilde\theta}F\,\csc(2\chi),
    \qquad
    Q=2\,\partial_{\tilde\theta}F\,\cot(2\chi).
\end{align}
For $F=\cos\tilde\theta$, this gives
\begin{align}
    \mathcal V_{(1)}
    =
    \cos\tilde\theta\,\partial_\chi
    -
    2\sin\tilde\theta\,\csc(2\chi)\partial_\phi
    -
    2\sin\tilde\theta\,\cot(2\chi)\td\theta.
\end{align}
For $F=\sin\tilde\theta$, this gives
\begin{align}
    \mathcal V_{(2)}
    =
    \sin\tilde\theta\,\partial_\chi
    +
    2\cos\tilde\theta\,\csc(2\chi)\partial_\phi
    +
    2\cos\tilde\theta\,\cot(2\chi)\td\theta.
\end{align}
Both satisfy
\begin{align}
    \widehat{\mathcal L}_{\mathcal V_{(i)}}\mathcal H=0,
    \qquad
    i=1,2.
\end{align}
They are not ordinary vector fields on the original $S^2\times S^1$ geometry, because they depend on the dual coordinate $\tilde\theta$ and contain a one-form component along $\td\theta$.

After T-duality,
\begin{align}
    \td\theta
    \longmapsto
    \partial_{\hat\theta},
    \qquad
    \tilde\theta
    \longmapsto
    \hat\theta.
\end{align}
Therefore the T-dual projections are
\begin{align}
    \mathcal P_\theta(\mathcal V_{(1)})
    &=
    \cos\hat\theta\,\partial_\chi
    -
    2\sin\hat\theta\,\csc(2\chi)\partial_\phi
    -
    2\sin\hat\theta\,\cot(2\chi)\partial_{\hat\theta},
    \\
    \mathcal P_\theta(\mathcal V_{(2)})
    &=
    \sin\hat\theta\,\partial_\chi
    +
    2\cos\hat\theta\,\csc(2\chi)\partial_\phi
    +
    2\cos\hat\theta\,\cot(2\chi)\partial_{\hat\theta}.
\end{align}
By the proposition, these are ordinary Killing vectors of the T-dual metric. In the dual co-frame
\begin{align}
    \hat e^1=\td\chi,
    \qquad
    \hat e^2=\frac12\sin(2\chi)\td\phi,
    \qquad
    \hat e^3=\frac12(\td\hat\theta-\cos(2\chi)\td\phi),
\end{align}
their metric-dual Killing 1-forms are
\begin{align}
    \hat K_{(1)}
    =
    \cos\hat\theta\,\hat e^1
    -
    \sin\hat\theta\,\hat e^2,
    \qquad
    \hat K_{(2)}
    =
    \sin\hat\theta\,\hat e^1
    +
    \cos\hat\theta\,\hat e^2.
\end{align}
These are precisely the two Killing 1-forms of the round $S^3$ which were not visible as ordinary Killing 1-forms of the original $S^2\times S^1$ geometry. Thus, in this example, the extra $S^3$ Killing 1-forms are obtained from the original generalized metric itself, not by solving the ordinary Killing equation on the T-dual metric.

\subsection{\texorpdfstring{$\phi$-dual Schwarzschild background}{phi-dual Schwarzschild background}}

The same mechanism appears in a four-dimensional example. Let
\begin{align}
    f(r)=1-\frac{2M}{r}.
\end{align}
Take the original background to be the $\phi$-dual Schwarzschild metric
\begin{align}
    \td s^2
    =
    -f(r)\td t^2
    +
    f(r)^{-1}\td r^2
    +
    r^2\td\theta^2
    +
    \frac{1}{r^2\sin^2\theta}\td\phi^2,
    \qquad
    B=0.
\end{align}
We apply T-duality along $y=\phi$. The T-dual metric is the ordinary Schwarzschild metric
\begin{align}
    \td\hat s^2
    =
    -f(r)\td t^2
    +
    f(r)^{-1}\td r^2
    +
    r^2\td\theta^2
    +
    r^2\sin^2\theta\,\td\hat\phi^2.
\end{align}
The original metric has only the manifest ordinary Killing vectors
\begin{align}
    \partial_t,
    \qquad
    \partial_\phi,
\end{align}
whereas the T-dual Schwarzschild metric has the full rotational $SO(3)$ algebra. As in the ordinary axial Buscher transformation, this construction is local on
the patch where the \(U(1)_\phi\) action is non-degenerate, namely
\(\sin\theta\neq 0\). The singular factors such as \(\cot\theta\) which appear
in the reconstructed generalized vectors are therefore understood as coordinate
patch effects associated with the polar axis.

We now recover the two non-axial rotations from the generalized Killing equation on the original background. For \(y=\phi\), the same ansatz-selection method gives generalized
vectors of the form
\begin{align}
    \mathcal V
    =
    v^\mu\partial_\mu+q\,\td\phi,
    \qquad
    \mu=t,r,\theta.
\end{align}
The additional Schwarzschild rotations are angular and do not involve
the \(t\)- or \(r\)-directions. Therefore, for the part relevant to the
two non-axial rotations, we use the reduced ansatz
\begin{align}
    \mathcal V
    =
    F(\theta,\tilde\phi)\partial_\theta
    +
    Q(\theta,\tilde\phi)\td\phi.
\end{align}

Substitution into $\widehat{\mathcal L}_{\mathcal V}\mathcal H=0$
gives
\begin{align}
    \partial_\theta F&=0, \\
    \partial_{\tilde\phi}Q&=-F\cot\theta, \\
    \partial_\theta Q
    &=
    -\csc^2\theta\,\partial_{\tilde\phi}F.
\end{align}
The compatibility condition is
\begin{align}
    \partial_{\tilde\phi}^2F+F=0.
\end{align}
Therefore the two real independent nonzero modes are
\begin{align}
    F=\sin\tilde\phi,
    \qquad
    F=\cos\tilde\phi.
\end{align}
For $F=\sin\tilde\phi$, one obtains
\begin{align}
    \mathcal V_1
    =
    \sin\tilde\phi\,\partial_\theta
    +
    \cot\theta\cos\tilde\phi\,\td\phi.
\end{align}
For $F=\cos\tilde\phi$, one obtains
\begin{align}
    \mathcal V_2
    =
    \cos\tilde\phi\,\partial_\theta
    -
    \cot\theta\sin\tilde\phi\,\td\phi.
\end{align}
Both satisfy
\begin{align}
    \widehat{\mathcal L}_{\mathcal V_i}\mathcal H=0,
    \qquad
    i=1,2.
\end{align}
They are not ordinary Killing vectors of the original metric because they contain a one-form component and depend on the dual coordinate $\tilde\phi$. After T-duality,
\begin{align}
    \td\phi
    \longmapsto
    \partial_{\hat\phi},
    \qquad
    \tilde\phi
    \longmapsto
    \hat\phi .
\end{align}
Therefore
\begin{align}
    \mathcal P_\phi(\mathcal V_1)
    &=
    \sin\hat\phi\,\partial_\theta
    +
    \cot\theta\cos\hat\phi\,\partial_{\hat\phi},
    \\
    \mathcal P_\phi(\mathcal V_2)
    &=
    \cos\hat\phi\,\partial_\theta
    -
    \cot\theta\sin\hat\phi\,\partial_{\hat\phi}.
\end{align}
By the proposition, these are ordinary Killing vectors of the T-dual Schwarzschild metric. Their metric-dual Killing 1-forms are
\begin{align}
    \hat K_1
    =
    r^2\sin\hat\phi\,\td\theta
    +
    r^2\sin\theta\cos\theta\cos\hat\phi\,\td\hat\phi,
\end{align}
and
\begin{align}
    \hat K_2
    =
    r^2\cos\hat\phi\,\td\theta
    -
    r^2\sin\theta\cos\theta\sin\hat\phi\,\td\hat\phi.
\end{align}
Equivalently, in the usual orthonormal co-frame
\begin{align}
    \hat e^\theta=r\,\td\theta,
    \qquad
    \hat e^\phi=r\sin\theta\,\td\hat\phi,
\end{align}
these are
\begin{align}
    \hat K_1
    =
    r\sin\hat\phi\,\hat e^\theta
    +
    r\cos\theta\cos\hat\phi\,\hat e^\phi,
\end{align}
and
\begin{align}
    \hat K_2
    =
    r\cos\hat\phi\,\hat e^\theta
    -
    r\cos\theta\sin\hat\phi\,\hat e^\phi.
\end{align}
These are the two non-axial rotational Killing 1-forms of the Schwarzschild spacetime. Thus the additional rotational symmetries of Schwarzschild are detected from the original $\phi$-dual generalized metric as non-geometric generalized Killing vectors.

\noindent The remaining two Killing vectors are also represented naturally in the
generalized tangent bundle. The time translation corresponds to the
purely vectorial generalized vector
\begin{align}
    \mathcal V_t=\partial_t+0,
\end{align}
whereas the axial rotation of the T-dual geometry is obtained from the
pure one-form generalized vector
\begin{align}
    \mathcal V_\phi=0+\td\phi .
\end{align}
Both satisfy the generalized Killing equation for the original
generalized metric,
\begin{align}
    \widehat{\mathcal L}_{\mathcal V_t}\mathcal H=0,
    \qquad
    \widehat{\mathcal L}_{\mathcal V_\phi}\mathcal H=0.
\end{align}
Under the T-dual projection,
\begin{align}
    \mathcal P_\phi(\mathcal V_t)
    &=
    \partial_t,
    &
    \mathcal P_\phi(\mathcal V_\phi)
    &=
    \partial_{\hat\phi}.
\end{align}
Together with \(\mathcal P_\phi(\mathcal V_1)\) and
\(\mathcal P_\phi(\mathcal V_2)\), these four vector fields span the
full vector space of ordinary Killing vectors of the T-dual
Schwarzschild metric.

The construction above applies directly to Killing 1-forms because they
are metric duals of Killing vectors. The relevant generalized-geometric
object is therefore the generalized Killing vector. The examples show
that emergent Killing 1-forms after T-duality can be detected before
duality by solving the generalized Killing equation for the original
generalized metric, provided one allows admissible dependence on the
dual coordinate associated with the T-dualized direction.

\noindent For higher-degree KY forms, the situation is different. A KY $p$-form for $p>1$ is not the metric dual of a vector field, and hence it is not directly captured by generalized Killing vectors. An analogous treatment would require an $O(d,d)$-covariant generalized KY equation whose projections reproduce the torsionful KY equations in different duality frames. The Killing 1-form construction above suggests the correct pattern. Emergent KY forms should correspond to generalized hidden-symmetry objects that are non-geometric in the original description but geometric after T-duality. 

\section{Conclusion} \label{sec6}

In this work we studied the behaviour of torsionful Killing--Yano forms under
Abelian T-duality. Using a coframe adapted to the isometry direction, we first
derived the transformation of the metric-compatible torsionful connection under
the Buscher rules. We then applied these formulae to the torsionful
Killing--Yano equation for a general \(p\)-form decomposed into transverse and
circle components.

The main result is a sufficient transformation criterion under explicit
matching assumptions on the auxiliary torsionful connections. Under these
assumptions, the original and T-dual Killing--Yano systems can be compared term
by term. The component of the Killing--Yano form along the dualized circle
transforms affinely, with an additive correction \(F\). This correction is not
arbitrary: it is constrained by two independent compatibility conditions,
\((F0)_a=0\) and \((F2)_a=0\), while \(F1=0\) arises as the exterior trace of
\((F2)_a=0\). We also recorded the restrictive special case \(A=b\), in which
the condition \(X_\theta(\alpha)=0\) can be relaxed at the price of modifying
the compatibility conditions on \(F\).

For \(p=1\), the general conditions reduce to a simple preservation statement.
A Killing 1-form is transformed into a Killing 1-form of the T-dual metric when
its transverse components are independent of the dualized isometry direction
and the scalar correction \(f\) satisfies the corresponding first-order
correction equation. Since the torsionful Killing--Yano equation for one-forms
is equivalent to the ordinary metric Killing equation for any totally
antisymmetric torsion, the one-form examples can be interpreted in terms of the
usual Killing symmetries of the metric.

The examples illustrate how these conditions distinguish between
Killing--Yano forms that are transformed by the direct T-duality map and those
that are not. In the Hopf T-dual of \(S^3\), only four of the six original
Killing 1-forms satisfy the required isometry-adapted condition and are mapped
to Killing 1-forms of the \(S^2\times S^1\) background. The Schwarzschild
examples show the same mechanism for dualities along different Killing
directions. The Nappi--Witten plane wave provides a torsionful example in which
the parallelizing NS--NS torsion allows the torsionful Killing--Yano 2-forms to
be transformed explicitly.

We also described the generalized-geometric origin of Killing 1-forms which
are present in the T-dual geometry but are not obtained from ordinary Killing
1-forms by the direct transformation map. In the \(S^2\times S^1\) and
\(\phi\)-dual Schwarzschild examples, the missing dual Killing vectors arise
from generalized Killing vectors of the original generalized metric. These
vectors are non-geometric in the original description, but become ordinary
Killing vector fields after the \(O(1,1)\) exchange associated with T-duality.
This supports the interpretation that T-duality may change which components of
a generalized symmetry are realized as ordinary geometric symmetries in a given
duality frame.

A natural direction for future work is to extend this generalized description
beyond Killing 1-forms. This would require a genuinely \(O(d,d)\)-covariant
formulation of higher-degree Killing--Yano structures whose projections
reproduce the torsionful Killing--Yano equations in different duality frames.
Such a formulation could clarify how hidden symmetries are encoded in
generalized geometry and how they transform under more general dualities,
including non-Abelian and Poisson--Lie T-duality.

\begin{acknowledgments}

This study was supported by the Scientific and Technological Research
Council of T\"urkiye (T\"UB\.ITAK) under Grant No.~123F261.
Ö.K. would like to thank Andrew Waldron for fruitful discussions.

\end{acknowledgments}

\appendix

\section{Metric-compatible torsionful connections}
\label{Appendix1}

We collect here the elementary facts about metric-compatible connections
with torsion that are used in Section~\ref{sec2}. Let \(\{e^a\}\) be a local
\(g\)-orthonormal co-frame with dual frame \(\{X_a\}\) \footnote{In this appendix only, the indices $a,b,c,\ldots$ denote generic
full-frame indices.}. A connection
\(\nabla\) is metric-compatible if its connection 1-forms
\(\omega^a{}_{b}\) satisfy
\begin{align}
    \nabla_{X}e^a
    =
    i_X(\omega^a{}_{b})\,e^b,
    \qquad
    \omega_{ab}=-\omega_{ba}.
\end{align}
The torsion 2-forms are defined by the first Cartan structure equation
\begin{align}
    T^a
    =
    \td e^a+\omega^a{}_{b}\wedge e^b .
\end{align}
Equivalently,
\begin{align}
    \td e^a-T^a
    =
    -\omega^a{}_{b}\wedge e^b .
\end{align}

For a fixed metric \(g\), a metric-compatible connection is uniquely
determined by its torsion tensor. Solving the first Cartan equation for
the connection 1-forms gives the standard expression~\cite{Benn_Tucker}:
\begin{align}
    \omega_{ab}
    =
    \frac12
    \bigg[
    i_{X_b}(\td e_a-T_a)
    -
    i_{X_a}(\td e_b-T_b)
    +
    i_{X_a}i_{X_b}(\td e_c-T_c)\,e^c
    \bigg].
    \label{appendix_connection_formula}
\end{align}
This is the formula used in Section~\ref{sec2} to compute the original and
T-dual torsionful connection components.

In the present paper, the torsion is assumed to be induced by a
totally antisymmetric 3-form. More precisely, we use the skew-torsion
3-form
\begin{align}
    \mathcal T=H+\Psi,
\end{align}
where \(H=\td B\) is the NS--NS flux and \(\Psi\) parametrizes the additional skew-torsion freedom of the auxiliary metric-compatible connection. The torsion 2-forms are then
\begin{align}
    T_a=i_{X_a}\mathcal T
    =
    i_{X_a}H+\psi_a,
    \qquad
    \psi_a:=i_{X_a}\Psi .
\end{align}
The same convention is used for the T-dual connection, with
\(\widehat{\mathcal T}=\widehat H+\widehat\Psi\).

Let
\begin{align}
    \Psi
    =
    \frac1{3!}\psi_{abc}\,e^a\wedge e^b\wedge e^c .
\end{align}
Then
\begin{align}
    \psi_a=i_{X_a}\Psi
    =
    \frac12\psi_{abc}\,e^b\wedge e^c .
\end{align}
Since \(\Psi\) is a 3-form, its components are totally antisymmetric:
\begin{align}
    \psi_{abc}
    =
    -\psi_{bac}
    =
    -\psi_{acb}.
\end{align}
Equivalently,
\begin{align}
    i_{X_a}\psi_b
    =
    -i_{X_b}\psi_a .
\end{align}
Conversely, if a collection of 2-forms \(\psi_a\) satisfies this
antisymmetry condition, then locally it arises from the 3-form
\begin{align}
    \Psi
    =
    \frac13\,e^a\wedge\psi_a .
\end{align}
Indeed,
\begin{align}
    i_{X_b}\Psi
    =
    \psi_b .
\end{align}
This justifies the use of the \(\psi_a\) and \(\hat\psi_a\) terms in
the torsion ansatz of Section~2.

For a torsionful metric-compatible connection \(\nabla^{\mathcal T}\),
the exterior derivative entering the torsionful Killing--Yano equation
is defined intrinsically by
\begin{align}
    \td^{\mathcal T}\omega
    :=
    e^a\wedge\nabla^{\mathcal T}_{X_a}\omega .
\end{align}
The torsionful Killing--Yano equation used in the main text is therefore
\begin{align}
    \nabla^{\mathcal T}_{X}\omega
    =
    \frac1{p+1}i_X\td^{\mathcal T}\omega .
\end{align}
In Sections~\ref{sec3} and~\ref{sec4} the superscript \(\mathcal T\) is suppressed for
notational simplicity, but the connection is always the
metric-compatible torsionful connection determined by the torsion
2-forms \(T_a=i_{X_a}(H+\Psi)\), or by their T-dual counterparts.

\section{Details of Computations}
\label{Appendix2}

\noindent This appendix presents the intermediate calculations leading
to the connection formulas in Section~\ref{sec2} and to the component
decompositions and matching conditions derived in Section~\ref{sec3}.
\subsection{Derivation of the torsionful connection relations}
\label{AppB_connections}

\noindent We apply the general connection formula
\eqref{appendix_connection_formula} to the original and T-dual
co-frames introduced in Section~\ref{sec2}. Splitting the full frame
index as $a'=(a,\theta)$, the transverse connection components of the
original geometry can be written as
\begin{align}\label{AppB_wab_split}
\omega_{ab}
={}&\frac12\bigg[
\underbrace{
 i_{X_b}(\td\te_a-T_a)
-i_{X_a}(\td\te_b-T_b)
+i_{X_a}i_{X_b}(\td\te_c-T_c)\te^c
}_{:=2\bar\omega_{ab}}
+i_{X_a}i_{X_b}(\td\te_\theta-T_\theta)\te^\theta
\bigg].
\end{align}
Here $\bar\omega_{ab}$ denotes the contribution obtained by restricting
the contracted index in \eqref{appendix_connection_formula} to the
transverse directions. Since
\begin{align}
\td\te^\theta
&=\td\sigma\wedge\te^\theta+e^\sigma\td A,
\nonumber\\
i_{X_a}i_{X_b}\td\te^\theta
&=e^\sigma i_{X_a}i_{X_b}\td A,
\qquad
i_{X_a}i_{X_b}T_\theta
=i_{X_a}i_{X_b}\bigl(e^{-\sigma}\td b+\psi_\theta\bigr),
\end{align}
we obtain
\begin{align}\label{AppB_wab_original}
\omega_{ab}
=
\bar\omega_{ab}
+\frac12 i_{X_a}i_{X_b}
\bigl(e^\sigma\td A-e^{-\sigma}\td b-\psi_\theta\bigr)
\te^\theta .
\end{align}

\noindent The mixed component follows from the same formula:
\begin{align}\label{AppB_wthetaa_original_raw}
\omega_{\theta a}
={}&\frac12\bigg[
 i_{X_a}(\td\te_\theta-T_\theta)
-i_{X_\theta}(\td\te_a-T_a)
+i_{X_\theta}i_{X_a}(\td\te_c-T_c)\te^c
+i_{X_\theta}i_{X_a}(\td\te_\theta-T_\theta)\te^\theta
\bigg]
\nonumber\\
={}&\frac12\bigg[
 i_{X_a}\td\te_\theta
-2i_{X_a}T_\theta
-\underbrace{i_{X_\theta}i_{X_a}T_c}_{i_{X_a}i_{X_c}T_\theta}\te^c
+(i_{X_\theta}i_{X_a}\td\te_\theta)\te^\theta
\bigg]
\nonumber\\
={}&\frac12\bigg[
(\td\sigma)_a\te^\theta
+e^\sigma i_{X_a}\td A
-2i_{X_a}\bigl(e^{-\sigma}\td b+\psi_\theta\bigr)
-i_{X_a}i_{X_c}\bigl(e^{-\sigma}\td b+\psi_\theta\bigr)\te^c
+(\td\sigma)_a\te^\theta
\bigg].
\end{align}
Consequently,
\begin{align}\label{AppB_wthetaa_original}
\omega_{\theta a}
=
(\td\sigma)_a\te^\theta
+\frac12 i_{X_a}
\bigl(e^\sigma\td A-e^{-\sigma}\td b-\psi_\theta\bigr).
\end{align}

\noindent To evaluate the transverse contribution in
\eqref{AppB_wab_split}, we use \eqref{eq:Ta} and the definition
\eqref{tau_def}. This gives
\begin{align}\label{AppB_barw_original}
\bar\omega_{ab}
={}&\frac12\bigg[
\underbrace{
 i_{X_b}\td\te_a-i_{X_a}\td\te_b
+\bigl(i_{X_a}i_{X_b}\td\te_c\bigr)\te^c
}_{2\tau_{ab}}
-i_{X_b}T_a+i_{X_a}T_b
-\bigl(i_{X_a}i_{X_b}T_c\bigr)\te^c
\bigg]
\nonumber\\
={}&\tau_{ab}
+i_{X_a}i_{X_b}
\bigl(h+e^{-\sigma}\td b\wedge\te^\theta\bigr)
+i_{X_a}\psi_b
-\frac12 i_{X_a}i_{X_b}
\bigl(i_{X_c}h+\psi_c\bigr)\te^c,
\end{align}
where $h$ is defined in \eqref{hlambda_def}.

\medskip

\noindent We now repeat the computation for the T-dual geometry. Since
the transverse co-frame is unchanged, $\he^a=\te^a$, one has
\begin{align}\label{AppB_whatab_split}
\hat\omega_{ab}
={}&\frac12\bigg[
\underbrace{
 i_{\hx_b}(\td\te_a-\hat T_a)
-i_{\hx_a}(\td\te_b-\hat T_b)
+i_{\hx_a}i_{\hx_b}(\td\te_c-\hat T_c)\te^c
}_{:=2\widehat{\bar\omega}_{ab}}
+i_{\hx_a}i_{\hx_b}
(\td\he_\theta-\hat T_\theta)\he^\theta
\bigg].
\end{align}
Using \eqref{eq:That_a} and \eqref{eq:That_theta}, the transverse part becomes
\begin{align}\label{AppB_barwhat}
\widehat{\bar\omega}_{ab}
={}&\tau_{ab}
+i_{\hx_a}i_{\hx_b}
\bigl(h+e^\sigma\td A\wedge\he^\theta\bigr)
+i_{\hx_a}\hp_b
-\frac12 i_{\hx_a}i_{\hx_b}
\bigl(i_{\hx_c}h+\hp_c\bigr)\te^c
\nonumber\\
={}&\bar\omega_{ab}
+\bigl(e^\sigma i_{X_a}i_{X_b}\td A-\hp_{b\theta a}\bigr)\he^\theta
-\bigl(e^{-\sigma}i_{X_a}i_{X_b}\td b-\psi_{b\theta a}\bigr)\te^\theta
+\frac12\bigl(\hp_{cba}-\psi_{cba}\bigr)\te^c.
\end{align}
Furthermore,
\begin{align}
\td\he^\theta
&=-\td\sigma\wedge\he^\theta+e^{-\sigma}\td b,
\nonumber\\
i_{\hx_a}i_{\hx_b}
(\td\he^\theta-\hat T_\theta)
&=i_{X_a}i_{X_b}
\bigl(e^{-\sigma}\td b-e^\sigma\td A-\hp_\theta\bigr).
\end{align}
It follows from \eqref{AppB_whatab_split} and
\eqref{AppB_barwhat} that
\begin{align}\label{AppB_whatab_intermediate}
\hat\omega_{ab}
={}&\widehat{\bar\omega}_{ab}
+\frac12 i_{\hx_a}i_{\hx_b}
\bigl(e^{-\sigma}\td b-e^\sigma\td A-\hp_\theta\bigr)
\he^\theta
\nonumber\\
={}&\bar\omega_{ab}
+\frac12 i_{X_a}i_{X_b}
\bigl(e^{-\sigma}\td b+e^\sigma\td A+\hp_\theta\bigr)
\he^\theta
\nonumber\\
&-i_{X_a}i_{X_b}
\bigl(e^{-\sigma}\td b+\psi_\theta\bigr)\te^\theta
+\frac12(\hp_{cba}-\psi_{cba})\te^c
\nonumber\\
={}&\omega_{ab}
+\frac12 i_{X_a}i_{X_b}
\bigl(e^{-\sigma}\td b+e^\sigma\td A+\hp_\theta\bigr)
\he^\theta
\nonumber\\
&-\frac12 i_{X_a}i_{X_b}
\bigl(e^\sigma\td A+e^{-\sigma}\td b+\psi_\theta\bigr)
\te^\theta
+\frac12(\hp_{cba}-\psi_{cba})\te^c.
\end{align}
Using the total antisymmetry of the components of $\psi_{a'}$ and
$\hp_{a'}$, this expression takes the compact form
\begin{align}\label{AppB_whatab_final}
\hat\omega_{ab}
={}&\omega_{ab}
+\frac12 i_{X_a}i_{X_b}
\bigl(e^{-\sigma}\td b+e^\sigma\td A\bigr)
(\he^\theta-\te^\theta)
+\frac12\bigl(i_{\hx_a}\hp_b-i_{X_a}\psi_b\bigr).
\end{align}

\noindent The mixed T-dual component is obtained in the same way:
\begin{align}\label{AppB_whathetaa}
\hat\omega_{\theta a}
={}&\frac12\bigg[
 i_{\hx_a}\td\he^\theta
-2i_{\hx_a}\hat T_\theta
-\bigl(i_{\hx_\theta}i_{\hx_a}\hat T_c\bigr)\te^c
+\bigl(i_{\hx_\theta}i_{\hx_a}\td\he_\theta\bigr)\he^\theta
\bigg]
\nonumber\\
={}&-(\td\sigma)_a\he^\theta
+\frac12 i_{X_a}
\bigl(e^{-\sigma}\td b-e^\sigma\td A-\hp_\theta\bigr)
\nonumber\\
={}&\omega_{\theta a}
-(\td\sigma)_a(\te^\theta+\he^\theta)
\nonumber\\
&\hspace{1.4cm}
+\frac12 i_{X_a}
\bigl(2e^{-\sigma}\td b-2e^\sigma\td A
+\psi_\theta-\hp_\theta\bigr).
\end{align}
Equations \eqref{AppB_whatab_final} and
\eqref{AppB_whathetaa} reproduce the connection one-form relations
stated in \eqref{wab_duals}. Throughout these expressions, the
differential forms are expanded in the common transverse orthonormal
co-frame; for example, $\td\sigma=(\td\sigma)_a\te^a$ and
$\td A=\frac12(\td A)_{ab}\te^a\wedge\te^b$.

\medskip

\noindent We next derive the corresponding covariant-derivative
relations. For either geometry,
\begin{align}\label{AppB_covariant_from_connection}
\nabla_{X_{a'}}\te_{b'}
=i_{X_{a'}}(\omega_{c'b'})\te^{c'}.
\end{align}
For the T-dual connection, it is convenient to separate the transverse
and circle components:
\begin{align}
\hat\nabla_{\hx_a}\he_b
&=i_{\hx_a}(\hat\omega_{cb})\te^c
+i_{\hx_a}(\hat\omega_{\theta b})\he^\theta,
\label{AppB_Tdconnection1}\\
\hat\nabla_{\hx_\theta}\he_a
&=i_{\hx_\theta}(\hat\omega_{ca})\te^c
+i_{\hx_\theta}(\hat\omega_{\theta a})\he^\theta,
\qquad
\hat\nabla_{\hx_a}\he_\theta
=i_{\hx_a}(\hat\omega_{b\theta})\te^b,
\label{AppB_Tdconnection2}\\
\hat\nabla_{\hx_\theta}\he_\theta
&=i_{\hx_\theta}(\hat\omega_{b\theta})\te^b.
\label{AppB_Tdconnection3}
\end{align}
Substituting \eqref{AppB_whatab_intermediate} and
\eqref{AppB_whathetaa} into \eqref{AppB_Tdconnection1} gives
\begin{align}\label{AppB_Tcon_ab}
\hat\nabla_{\hx_a}\he_b
={}&i_{\hx_a}(\bar\omega_{cb})\te^c
-(i_{\hx_a}\te^\theta)
 i_{X_c}i_{X_b}\bigl(e^{-\sigma}\td b+\psi_\theta\bigr)\te^c
\nonumber\\
&+\frac12(\hp_{abc}-\psi_{abc})\te^c
+\frac12 i_{\hx_a}i_{X_b}
\bigl(e^{-\sigma}\td b-e^\sigma\td A-\hp_\theta\bigr)
\he^\theta
\nonumber\\
={}&(i_{X_a}\tau_{cb})\te^c
-\frac12 i_{X_a}i_{X_b}h
+f_{ab}\he^\theta
+\frac12\hp_{\theta ab}\he^\theta
+\frac12\hp_{abc}\te^c.
\end{align}
The corresponding original component is
\begin{align}\label{AppB_con_ab}
\nabla_{X_a}\te_b
={}&i_{X_a}(\bar\omega_{cb})\te^c
+i_{X_a}(\omega_{\theta b})\te^\theta
\nonumber\\
={}&(i_{X_a}\tau_{cb})\te^c
-\frac12 i_{X_a}i_{X_b}h
-f_{ab}\te^\theta
+\frac12\psi_{\theta ab}\te^\theta
+\frac12\psi_{abc}\te^c.
\end{align}
Subtracting \eqref{AppB_con_ab} from \eqref{AppB_Tcon_ab}, and using
the antisymmetry of the torsion components, yields
\begin{align}\label{AppB_cov_ab_relation}
\hat\nabla_{\hx_a}\he_b
=
\nabla_{X_a}\te_b
+f_{ab}(\te^\theta+\he^\theta)
+\frac12\bigl(i_{\hx_b}\hp_a-i_{X_b}\psi_a\bigr).
\end{align}

\noindent The remaining components follow directly from
\eqref{AppB_wthetaa_original} and \eqref{AppB_whathetaa}:
\begin{align}\label{AppB_connection_components}
\hat\nabla_{\hx_\theta}\he_a
&=
\frac12 i_{X_a}
\bigl(e^{-\sigma}\td b+e^\sigma\td A+\hp_\theta\bigr)
-(\td\sigma)_a\he^\theta,
\nonumber\\
\nabla_{X_\theta}\te_a
&=
\frac12 i_{X_a}
\bigl(e^\sigma\td A+e^{-\sigma}\td b+\psi_\theta\bigr)
+(\td\sigma)_a\te^\theta,
\nonumber\\
\hat\nabla_{\hx_a}\he_\theta
&=f_{ab}\te^b-\frac12 i_{X_a}\hp_\theta,
\qquad
\nabla_{X_a}\te_\theta
=-f_{ab}\te^b-\frac12 i_{X_a}\psi_\theta,
\nonumber\\
\hat\nabla_{\hx_\theta}\he_\theta
&=\td\sigma,
\qquad
\nabla_{X_\theta}\te_\theta
=-\td\sigma.
\end{align}
Therefore,
\begin{align}\label{AppB_covariant_relations_final}
\hat\nabla_{\hx_a}\he_b
&=
\nabla_{X_a}\te_b
+f_{ab}(\te^\theta+\he^\theta)
+\frac12\bigl(i_{\hx_b}\hp_a-i_{X_b}\psi_a\bigr),
\nonumber\\
\hat\nabla_{\hx_\theta}\he_a
&=
\nabla_{X_\theta}\te_a
-(\td\sigma)_a(\te^\theta+\he^\theta)
+\frac12 i_{X_a}(\hp_\theta-\psi_\theta),
\nonumber\\
\hat\nabla_{\hx_a}\he_\theta
&=
-\nabla_{X_a}\te_\theta
-\frac12 i_{X_a}(\hp_\theta+\psi_\theta),
\nonumber\\
\hat\nabla_{\hx_\theta}\he_\theta
&=-\nabla_{X_\theta}\te_\theta.
\end{align}
As a consistency check, the original and dual component formulas are
interchanged by $\sigma\leftrightarrow-\sigma$ and
$b\leftrightarrow A$, together with the exchange of hatted and
unhatted quantities. Equations \eqref{AppB_covariant_relations_final}
are the covariant-derivative relations stated in Section~\ref{sec2}
and used in Section~\ref{sec3}.
\subsection{Component decomposition of the original and T-dual
Killing--Yano equations}
\label{AppB_KY_decomposition}

\noindent We begin with the algebraic form of the torsionful
Killing--Yano equation,
\begin{align}
0
=
p\,\nabla_{X_{a'}}K
+
e^{b'}\wedge i_{X_{a'}}
\bigl(\nabla_{X_{b'}}K\bigr).
\end{align}
Splitting the full frame index as $a'=\{a,\theta\}$ gives
\begin{align}\label{AppB_KY_split}
0
={}&
p\,\nabla_{X_a}K
+
e^b\wedge i_{X_a}\bigl(\nabla_{X_b}K\bigr)
+
e^\theta\wedge i_{X_a}
\bigl(\nabla_{X_\theta}K\bigr),
\\
0
={}&
p\,\nabla_{X_\theta}K
+
e^b\wedge i_{X_\theta}\bigl(\nabla_{X_b}K\bigr)
+
e^\theta\wedge i_{X_\theta}
\bigl(\nabla_{X_\theta}K\bigr).
\nonumber
\end{align}

\noindent We now substitute
\begin{align}
K=\alpha+e^\theta\wedge\beta,
\end{align}
where $\alpha$ is a transverse $p$-form and $\beta$ is a transverse
$(p-1)$-form. The $X_a$ equation in \eqref{AppB_KY_split} becomes
\begin{align}\label{AppB_KY1a}
0={}&
p\Big(
\nabla_{X_a}\alpha
+
(\nabla_{X_a}e^\theta)\wedge\beta
+
e^\theta\wedge\nabla_{X_a}\beta
\Big)
\nonumber\\
&+
e^b\wedge
\Big(
i_{X_a}\nabla_{X_b}\alpha
+
i_{X_a}(\nabla_{X_b}e^\theta)\wedge\beta
-
(\nabla_{X_b}e^\theta)\wedge i_{X_a}\beta
-
e^\theta\wedge i_{X_a}\nabla_{X_b}\beta
\Big)
\nonumber\\
&+
e^\theta\wedge
\Big(
i_{X_a}\nabla_{X_\theta}\alpha
+
i_{X_a}(\nabla_{X_\theta}e^\theta)\wedge\beta
-
(\nabla_{X_\theta}e^\theta)\wedge i_{X_a}\beta
\Big),
\end{align}
whereas the $X_\theta$ equation gives
\begin{align}\label{AppB_KY1theta}
0={}&
p\Big(
\nabla_{X_\theta}\alpha
+
(\nabla_{X_\theta}e^\theta)\wedge\beta
+
e^\theta\wedge\nabla_{X_\theta}\beta
\Big)
\nonumber\\
&+
e^b\wedge
\Big(
i_{X_\theta}\nabla_{X_b}\alpha
+
i_{X_\theta}(\nabla_{X_b}e^\theta)\wedge\beta
+
\nabla_{X_b}\beta
-
e^\theta\wedge i_{X_\theta}\nabla_{X_b}\beta
\Big)
\nonumber\\
&+
e^\theta\wedge
\Big(
i_{X_\theta}\nabla_{X_\theta}\alpha
+
i_{X_\theta}(\nabla_{X_\theta}e^\theta)\wedge\beta
+
\nabla_{X_\theta}\beta
\Big).
\end{align}

\noindent Inserting the explicit covariant derivatives derived in
Subsection~\ref{AppB_connections} into
\eqref{AppB_KY1a} and \eqref{AppB_KY1theta}, we obtain
\begin{align}\label{AppB_KY2a}
0={}&
p\Big(
\nabla_{X_a}\alpha
-
\bigl(f_{ak}e^k+\tfrac12 i_{X_a}\psi_\theta\bigr)
 \wedge\beta
+
e^\theta\wedge\nabla_{X_a}\beta
\Big)
\nonumber\\
&+
e^b\wedge
\Big(
i_{X_a}\nabla_{X_b}\alpha
-
\bigl(f_{ba}+\tfrac12 i_{X_a}i_{X_b}\psi_\theta\bigr)\beta
+
\bigl(f_{bk}e^k+\tfrac12 i_{X_b}\psi_\theta\bigr)
 \wedge i_{X_a}\beta
-
e^\theta\wedge i_{X_a}\nabla_{X_b}\beta
\Big)
\nonumber\\
&+
e^\theta\wedge
\Big(
i_{X_a}\nabla_{X_\theta}\alpha
-
(i_{X_a}d\sigma)\beta
+
d\sigma\wedge i_{X_a}\beta
\Big),
\end{align}
and
\begin{align}\label{AppB_KY2theta}
0={}&
p\Big(
\nabla_{X_\theta}\alpha
-
d\sigma\wedge\beta
+
e^\theta\wedge\nabla_{X_\theta}\beta
\Big)
\nonumber\\
&+
e^b\wedge
\Big(
i_{X_\theta}\nabla_{X_b}\alpha
+
\nabla_{X_b}\beta
-
e^\theta\wedge i_{X_\theta}\nabla_{X_b}\beta
\Big)+
e^\theta\wedge
\Big(
i_{X_\theta}\nabla_{X_\theta}\alpha
+
\nabla_{X_\theta}\beta
\Big).
\end{align}

\noindent For any transverse form $\gamma$, we use
\begin{align}\label{AppB_form_derivative}
\nabla_{X_{a'}}\gamma
=
X_{a'}(\gamma)
+
(\nabla_{X_{a'}}e^m)\wedge i_{X_m}\gamma.
\end{align}
Applying this identity to $\alpha$ and $\beta$ in
\eqref{AppB_KY2a} gives
\begin{adjustwidth}{-1.8cm}{-1.8cm}
\begin{align}\label{AppB_KY3a}
0={}&
p\Big(
X_a(\alpha)
+
(\nabla_{X_a}e^m)\wedge i_{X_m}\alpha
-
\bigl(f_{ak}e^k+\tfrac12 i_{X_a}\psi_\theta\bigr)
 \wedge\beta
+
e^\theta\wedge
\bigl[
X_a(\beta)
+
(\nabla_{X_a}e^m)\wedge i_{X_m}\beta
\bigr]
\Big)
\nonumber\\
&+
e^b\wedge
\Big(
i_{X_a}
\bigl[
X_b(\alpha)
+
(\nabla_{X_b}e^m)\wedge i_{X_m}\alpha
\bigr]
-
\bigl(f_{ba}+\tfrac12 i_{X_a}i_{X_b}\psi_\theta\bigr)
 \beta
+
\bigl(f_{bk}e^k+\tfrac12 i_{X_b}\psi_\theta\bigr)
 \wedge i_{X_a}\beta
\Big)
\nonumber\\
&+
e^\theta\wedge
\Big(
e^b\wedge i_{X_a}
\bigl[
X_b(\beta)
+
(\nabla_{X_b}e^m)\wedge i_{X_m}\beta
\bigr]
+
i_{X_a}
\bigl[
X_\theta(\alpha)
+
(\nabla_{X_\theta}e^m)\wedge i_{X_m}\alpha
\bigr]
-
(i_{X_a}d\sigma)\beta
+
d\sigma\wedge i_{X_a}\beta
\Big).
\end{align}

\noindent Similarly, \eqref{AppB_KY2theta} becomes
\begin{align}\label{AppB_KY3theta}
0={}&
p\Big(
X_\theta(\alpha)
+
(\nabla_{X_\theta}e^m)\wedge i_{X_m}\alpha
-
d\sigma\wedge\beta
+
e^\theta\wedge
\bigl[
X_\theta(\beta)
+
(\nabla_{X_\theta}e^m)\wedge i_{X_m}\beta
\bigr]
\Big)
\nonumber\\
&+
e^b\wedge
\Big(
i_{X_\theta}
\bigl[
X_b(\alpha)
+
(\nabla_{X_b}e^m)\wedge i_{X_m}\alpha
\bigr]
+
X_b(\beta)
+
(\nabla_{X_b}e^m)\wedge i_{X_m}\beta
\nonumber\\
&\hspace{2.7cm}
-
e^\theta\wedge i_{X_\theta}
\bigl[
X_b(\beta)
+
(\nabla_{X_b}e^m)\wedge i_{X_m}\beta
\bigr]
\Big)
\nonumber\\
&+
e^\theta\wedge
\Big(
i_{X_\theta}
\bigl[
X_\theta(\alpha)
+
(\nabla_{X_\theta}e^m)\wedge i_{X_m}\alpha
\bigr]
+
X_\theta(\beta)
+
(\nabla_{X_\theta}e^m)\wedge i_{X_m}\beta
\Big).
\end{align}
\end{adjustwidth}

\noindent Substituting the explicit expression for
$\nabla_{X_a}e^m$ into \eqref{AppB_KY3a} yields
\begin{adjustwidth}{-1.8cm}{-1.8cm}
\begin{align}\label{AppB_KY4a}
0={}&
p\Big(
X_a(\alpha)
+
\bigl(
i_{X_a}\tau_c{}^m
+\tfrac12\psi_a{}^m{}_c
\bigr)e^c\wedge i_{X_m}\alpha
-
\tfrac12
\bigl(i_{X_a}i_{X^m}h\bigr)\wedge i_{X_m}\alpha
\nonumber\\
&\qquad
-
\bigl(
f_a{}^m-\tfrac12\psi_{\theta a}{}^m
\bigr)e^\theta\wedge i_{X_m}\alpha
-
\bigl(
f_{ak}e^k+\tfrac12 i_{X_a}\psi_\theta
\bigr)\wedge\beta
\nonumber\\
&\qquad
+
e^\theta\wedge
\Big[
X_a(\beta)
+
\bigl(
i_{X_a}\tau_c{}^m
+\tfrac12\psi_a{}^m{}_c
\bigr)e^c\wedge i_{X_m}\beta
-
\tfrac12
\bigl(i_{X_a}i_{X^m}h\bigr)\wedge i_{X_m}\beta
\Big]
\Big)
\nonumber\\
&+
e^b\wedge
\Big(
i_{X_a}X_b(\alpha)
+
\bigl(
i_{X_b}\tau_a{}^m
+\tfrac12\psi_b{}^m{}_a
\bigr)i_{X_m}\alpha
\nonumber\\
&\qquad\qquad
-
\bigl(
i_{X_b}\tau_c{}^m
+\tfrac12\psi_b{}^m{}_c
\bigr)e^c\wedge i_{X_a}i_{X_m}\alpha
-
\tfrac12 i_{X_a}
\Big[
\bigl(i_{X_b}i_{X^m}h\bigr)\wedge i_{X_m}\alpha
\Big]
\nonumber\\
&\qquad\qquad
+
\bigl(
f_b{}^m-\tfrac12\psi_{\theta b}{}^m
\bigr)e^\theta\wedge i_{X_a}i_{X_m}\alpha
-
\bigl(
f_{ba}+\tfrac12 i_{X_a}i_{X_b}\psi_\theta
\bigr)\beta
\nonumber\\
&\qquad\qquad
+
\bigl(
f_{bk}e^k+\tfrac12 i_{X_b}\psi_\theta
\bigr)\wedge i_{X_a}\beta
\Big)
\nonumber\\
&+
e^\theta\wedge
\Big(
e^b\wedge i_{X_a}X_b(\beta)
\nonumber\\
&\qquad
+
e^b\wedge
\Big[
\bigl(
i_{X_b}\tau_a{}^m
+\tfrac12\psi_b{}^m{}_a
\bigr)i_{X_m}\beta
-
\bigl(
i_{X_b}\tau_c{}^m
+\tfrac12\psi_b{}^m{}_c
\bigr)e^c\wedge i_{X_a}i_{X_m}\beta
\nonumber\\
&\qquad\qquad
-
\tfrac12 i_{X_a}
\Big[
\bigl(i_{X_b}i_{X^m}h\bigr)\wedge i_{X_m}\beta
\Big]
\Big]
+
i_{X_a}X_\theta(\alpha)
\nonumber\\
&\qquad
+
\tfrac12 i_{X_a}
\Big[
i_{X^m}(\Lambda+\psi_\theta)
\wedge i_{X_m}\alpha
\Big]
-
(i_{X_a}d\sigma)\beta
+
d\sigma\wedge i_{X_a}\beta
\Big).
\end{align}
\end{adjustwidth}

\noindent Separating \eqref{AppB_KY4a} into its transverse part and
its $e^\theta$-mixed part gives precisely
\eqref{KYorga1} and \eqref{KYorga2}, respectively. Applying the same
substitution to \eqref{AppB_KY3theta} and separating the two parts
gives \eqref{KYorgth1} and \eqref{KYorgth2}. These are the four
original component equations displayed in Section~\ref{sec3}.

\medskip

\noindent We next repeat the decomposition for the T-dual form
\begin{align}
\hat K=\hat\alpha+\he^\theta\wedge\hat\beta.
\end{align}
Using the hatted covariant derivatives obtained in
Subsection~\ref{AppB_connections}, the transverse and mixed parts of
the $\hx_a$ and $\hx_\theta$ equations give the following four
intermediate equations:
\begin{adjustwidth}{-1.8cm}{-1.8cm}
\begin{equation}\label{AppB_dual1}
\begin{aligned}
0={}&
p\Big(
\hx_a(\hat\alpha)
+
\Big(
i_{X_a}\tau_c{}^m
+\frac12\hat\psi_a{}^m{}_c
\Big)e^c\wedge i_{X_m}\hat\alpha
-
\frac12
\bigl(i_{X_a}i_{X^m}h\bigr)
\wedge i_{X_m}\hat\alpha
\\
&\hspace{2.2cm}
+
\Big(
f_{ak}e^k-\frac12 i_{X_a}\hat\psi_\theta
\Big)\wedge\hat\beta
\Big)
\\
&\quad
+
e^b\wedge
\Big(
i_{X_a}\hx_b(\hat\alpha)
+
\Big(
i_{X_b}\tau_a{}^m
+\frac12\hat\psi_b{}^m{}_a
\Big)i_{X_m}\hat\alpha
\\
&\hspace{2.2cm}
-
\Big(
i_{X_b}\tau_c{}^m
+\frac12\hat\psi_b{}^m{}_c
\Big)e^c\wedge i_{X_a}i_{X_m}\hat\alpha
-
\frac12 i_{X_a}
\Big[
\bigl(i_{X_b}i_{X^m}h\bigr)
\wedge i_{X_m}\hat\alpha
\Big]
\\
&\hspace{2.2cm}
+
\Big(
f_{ba}-\frac12 i_{X_a}i_{X_b}\hat\psi_\theta
\Big)\hat\beta
-
\Big(
f_{bk}e^k-\frac12 i_{X_b}\hat\psi_\theta
\Big)\wedge i_{X_a}\hat\beta
\Big).
\end{aligned}
\end{equation}

\begin{equation}\label{AppB_dual2}
\begin{aligned}
0={}&
p\Big(
\hx_a(\hat\beta)
+
\Big(
i_{X_a}\tau_c{}^m
+\frac12\hat\psi_a{}^m{}_c
\Big)e^c\wedge i_{X_m}\hat\beta
-
\frac12
\bigl(i_{X_a}i_{X^m}h\bigr)
\wedge i_{X_m}\hat\beta
\\
&\hspace{2.2cm}
+
\Big(
f_a{}^m+\frac12\hat\psi_{\theta a}{}^m
\Big)i_{X_m}\hat\alpha
\Big)
\\
&\quad
+
e^b\wedge
\Big(
i_{X_a}\hx_b(\hat\beta)
+
\Big(
i_{X_b}\tau_a{}^m
+\frac12\hat\psi_b{}^m{}_a
\Big)i_{X_m}\hat\beta
\\
&\hspace{2.2cm}
-
\Big(
i_{X_b}\tau_c{}^m
+\frac12\hat\psi_b{}^m{}_c
\Big)e^c\wedge i_{X_a}i_{X_m}\hat\beta
-
\frac12 i_{X_a}
\Big[
\bigl(i_{X_b}i_{X^m}h\bigr)
\wedge i_{X_m}\hat\beta
\Big]
\\
&\hspace{2.2cm}
+
\Big(
f_b{}^m+\frac12\hat\psi_{\theta b}{}^m
\Big)i_{X_a}i_{X_m}\hat\alpha
\Big)
\\
&\quad
+
i_{X_a}\hx_\theta(\hat\alpha)
+
\frac12 i_{X_a}
\Big[
i_{X^m}(\Lambda+\hat\psi_\theta)
\wedge i_{X_m}\hat\alpha
\Big]
\\
&\quad
+
(i_{X_a}d\sigma)\hat\beta
-
d\sigma\wedge i_{X_a}\hat\beta .
\end{aligned}
\end{equation}

\begin{equation}\label{AppB_dual3}
\begin{aligned}
0={}&
p\,\hx_\theta(\hat\alpha)
+
p\,d\sigma\wedge\hat\beta
\\
&\quad
+
i_{X^m}
\Big[
\frac{p+1}{2}e^{-\sigma}db
+
\frac{p-1}{2}
\bigl(\hat\psi_\theta+e^\sigma dA\bigr)
\Big]\wedge i_{X_m}\hat\alpha
\\
&\quad
+
e^b\wedge\hx_b(\hat\beta)
+
\Big(
\tau_c{}^m\wedge e^c
-
i_{X^m}h
-
\frac12\hat\psi^m{}_{bc}\,
e^b\wedge e^c
\Big)\wedge i_{X_m}\hat\beta .
\end{aligned}
\end{equation}

\begin{equation}\label{AppB_dual4}
0
=
(p+1)
\Big[
\hx_\theta(\hat\beta)
-
(i_{X^m}d\sigma)i_{X_m}\hat\alpha
+
\frac12 i_{X^m}
(\Lambda+\hat\psi_\theta)
\wedge i_{X_m}\hat\beta
\Big].
\end{equation}
\end{adjustwidth}

\noindent Finally, substituting the transformed frame vectors
\eqref{basisvecs} into \eqref{AppB_dual1}--\eqref{AppB_dual4} gives
\eqref{3.20}--\eqref{3.23}. For example,
\begin{align}
\hx_a(\hat\alpha)
&=
X_a(\hat\alpha)
+
e^\sigma i_{X_a}(A-b)\,
X_\theta(\hat\alpha),
\nonumber\\
e^b\wedge i_{X_a}\hx_b(\hat\alpha)
&=
e^b\wedge i_{X_a}X_b(\hat\alpha)
+
e^\sigma(A-b)\wedge
i_{X_a}X_\theta(\hat\alpha),
\nonumber\\
\hx_\theta(\hat\alpha)
&=
e^{2\sigma}X_\theta(\hat\alpha),
\end{align}
and the same substitutions apply to $\hat\beta$. This completes the
derivation of the original and T-dual component systems stated in Section~\ref{sec3}.

\subsection{Reduction under the matching assumptions}
\label{AppB_matching}

\noindent We now impose the matching assumptions
\eqref{matching_conditions} on the original system
\eqref{KYorga1}--\eqref{KYorgth2} and the T-dual system
\eqref{3.20}--\eqref{3.23}. The original equations reduce to
\begin{adjustwidth}{-1.8cm}{-1.8cm}
\begin{equation}\label{AppB_matched_original1}
\begin{aligned}
0={}&
p\Big(
X_a(\alpha)
-
\bigl(
f_{ak}e^k-\tfrac12 i_{X_a}\Lambda
\bigr)\wedge\beta
\Big)
\\
&\quad
+
e^b\wedge
\Big(
i_{X_a}X_b(\alpha)
-
\bigl(
f_{ba}-\tfrac12 i_{X_a}i_{X_b}\Lambda
\bigr)\beta
\\
&\hspace{3.2cm}
+
\bigl(
f_{bk}e^k-\tfrac12 i_{X_b}\Lambda
\bigr)\wedge i_{X_a}\beta
\Big).
\end{aligned}
\end{equation}

\begin{equation}\label{AppB_matched_original2}
\begin{aligned}
0={}&
p\Big(
X_a(\beta)
-
\bigl(
f_a{}^m+\tfrac12 i_{X^m}i_{X_a}\Lambda
\bigr)i_{X_m}\alpha
\Big)
\\
&\quad
+
e^b\wedge
\Big(
i_{X_a}X_b(\beta)
-
\bigl(
f_b{}^m+\tfrac12 i_{X^m}i_{X_b}\Lambda
\bigr)i_{X_a}i_{X_m}\alpha
\Big)
\\
&\quad
-
(i_{X_a}d\sigma)\beta
+
d\sigma\wedge i_{X_a}\beta .
\end{aligned}
\end{equation}

\begin{equation}\label{AppB_matched_original3}
0
=
-p\,d\sigma\wedge\beta
+
i_{X^m}(e^\sigma dA)
\wedge i_{X_m}\alpha
+
e^b\wedge X_b(\beta).
\end{equation}

\begin{equation}\label{AppB_matched_original4}
0
=
X_\theta(\beta)
+
(i_{X^m}d\sigma)i_{X_m}\alpha .
\end{equation}
\end{adjustwidth}

\noindent Under the same assumptions, the T-dual equations become
\begin{adjustwidth}{-1.8cm}{-1.8cm}
\begin{equation}\label{AppB_matched_dual1}
\begin{aligned}
0={}&
p\Big(
X_a(\alpha)
+
\bigl(
f_{ak}e^k+\tfrac12 i_{X_a}\Lambda
\bigr)\wedge\hat\beta
\Big)
\\
&\quad
+
e^b\wedge
\Big(
i_{X_a}X_b(\alpha)
+
\bigl(
f_{ba}+\tfrac12 i_{X_a}i_{X_b}\Lambda
\bigr)\hat\beta
\\
&\hspace{3.2cm}
-
\bigl(
f_{bk}e^k+\tfrac12 i_{X_b}\Lambda
\bigr)\wedge i_{X_a}\hat\beta
\Big).
\end{aligned}
\end{equation}

\begin{equation}\label{AppB_matched_dual2}
\begin{aligned}
0={}&
p\Big(
X_a(\hat\beta)
+
e^\sigma i_{X_a}(A-b)\,
X_\theta(\hat\beta)
+
\bigl(
f_a{}^m-\tfrac12 i_{X^m}i_{X_a}\Lambda
\bigr)i_{X_m}\alpha
\Big)
\\
&\quad
+
e^b\wedge
\Big(
i_{X_a}X_b(\hat\beta)
+
\bigl(
f_b{}^m-\tfrac12 i_{X^m}i_{X_b}\Lambda
\bigr)i_{X_a}i_{X_m}\alpha
\Big)
\\
&\quad
+
e^\sigma(A-b)\wedge
i_{X_a}X_\theta(\hat\beta)
+
(i_{X_a}d\sigma)\hat\beta
-
d\sigma\wedge i_{X_a}\hat\beta .
\end{aligned}
\end{equation}

\begin{equation}\label{AppB_matched_dual3}
\begin{aligned}
0={}&
p\,d\sigma\wedge\hat\beta
+
i_{X^m}(e^{-\sigma}db)
\wedge i_{X_m}\alpha
+
e^b\wedge X_b(\hat\beta)
\\
&\quad
+
e^\sigma(A-b)\wedge X_\theta(\hat\beta).
\end{aligned}
\end{equation}

\begin{equation}\label{AppB_matched_dual4}
0
=
e^{2\sigma}X_\theta(\hat\beta)
-
(i_{X^m}d\sigma)i_{X_m}\alpha .
\end{equation}
\end{adjustwidth}

\noindent The opposite signs of the circle-scale terms in
\eqref{AppB_matched_original4} and \eqref{AppB_matched_dual4}
motivate the affine transformation ansatz \eqref{bsol}. Using the
definitions \eqref{hlambda_def} and \eqref{fab_def}, the original
system takes the form

\begin{adjustwidth}{-1.8cm}{-1.8cm}
\begin{equation}\label{AppB_reduced_original1}
\begin{aligned}
0={}&
p\Big(
X_a(\alpha)
+
e^{-\sigma}i_{X_a}db\wedge\beta
\Big)
\\
&\quad
+
e^b\wedge
\Big(
i_{X_a}X_b(\alpha)
+
e^{-\sigma}
(i_{X_a}i_{X_b}db)\beta
-
e^{-\sigma}i_{X_b}db
\wedge i_{X_a}\beta
\Big).
\end{aligned}
\end{equation}

\begin{equation}\label{AppB_reduced_original2}
\begin{aligned}
0={}&
p\Big(
X_a(\beta)
+
e^\sigma i_{X_a}i_{X^m}dA\,
i_{X_m}\alpha
\Big)
\\
&\quad
+
e^b\wedge
\Big(
i_{X_a}X_b(\beta)
+
e^\sigma i_{X_b}i_{X^m}dA\,
i_{X_a}i_{X_m}\alpha
\Big)
\\
&\quad
-
(i_{X_a}d\sigma)\beta
+
d\sigma\wedge i_{X_a}\beta .
\end{aligned}
\end{equation}

\begin{equation}\label{AppB_reduced_original3}
0
=
-p\,d\sigma\wedge\beta
+
e^\sigma i_{X^m}dA
\wedge i_{X_m}\alpha
+
e^b\wedge X_b(\beta).
\end{equation}

\begin{equation}\label{AppB_reduced_original4}
0
=
X_\theta(\beta)
+
(i_{X^m}d\sigma)i_{X_m}\alpha .
\end{equation}
\end{adjustwidth}

\noindent Substituting
\[
\hat\beta=-e^{-2\sigma}\beta+F,
\qquad
X_\theta(F)=0,
\]
into the T-dual system gives

\begin{adjustwidth}{-1.8cm}{-1.8cm}
\begin{equation}\label{AppB_reduced_dual1}
\begin{aligned}
0={}&
p\Big(
X_a(\alpha)
-
e^{-\sigma}i_{X_a}dA\wedge\beta
+
e^\sigma i_{X_a}dA\wedge F
\Big)
\\
&\quad
+
e^b\wedge
\Big(
i_{X_a}X_b(\alpha)
-
e^{-\sigma}
(i_{X_a}i_{X_b}dA)\beta
+
e^\sigma
(i_{X_a}i_{X_b}dA)F
\\
&\hspace{3.2cm}
+
e^{-\sigma}i_{X_b}dA
\wedge i_{X_a}\beta
-
e^\sigma i_{X_b}dA
\wedge i_{X_a}F
\Big).
\end{aligned}
\end{equation}

\begin{equation}\label{AppB_reduced_dual2}
\begin{aligned}
0={}&
p\Big(
X_a(F)
-
e^{-2\sigma}X_a(\beta)
+
2e^{-2\sigma}
(i_{X_a}d\sigma)\beta
-
e^{-\sigma}i_{X_a}(A-b)\,
X_\theta(\beta)
\\
&\hspace{2.2cm}
+
e^{-\sigma}
i_{X_a}i_{X^m}db\,
i_{X_m}\alpha
\Big)
\\
&\quad
+
e^b\wedge
\Big(
i_{X_a}X_b(F)
-
e^{-2\sigma}i_{X_a}X_b(\beta)
+
2e^{-2\sigma}
(i_{X_b}d\sigma)i_{X_a}\beta
\\
&\hspace{2.2cm}
+
e^{-\sigma}
i_{X_b}i_{X^m}db\,
i_{X_a}i_{X_m}\alpha
\Big)
\\
&\quad
-
e^{-\sigma}(A-b)\wedge
i_{X_a}X_\theta(\beta)
-
e^{-2\sigma}
(i_{X_a}d\sigma)\beta
+
e^{-2\sigma}d\sigma\wedge i_{X_a}\beta
\\
&\quad
+
(i_{X_a}d\sigma)F
-
d\sigma\wedge i_{X_a}F .
\end{aligned}
\end{equation}

\begin{equation}\label{AppB_reduced_dual3}
\begin{aligned}
0={}&
p\,d\sigma\wedge F
+
(2-p)e^{-2\sigma}d\sigma\wedge\beta
+
e^{-\sigma}i_{X^m}db
\wedge i_{X_m}\alpha
\\
&\quad
+
e^b\wedge X_b(F)
-
e^{-2\sigma}e^b\wedge X_b(\beta)
-
e^{-\sigma}(A-b)\wedge X_\theta(\beta).
\end{aligned}
\end{equation}

\begin{equation}\label{AppB_reduced_dual4}
0
=
-X_\theta(\beta)
-
(i_{X^m}d\sigma)i_{X_m}\alpha .
\end{equation}
\end{adjustwidth}

\noindent Equation \eqref{AppB_reduced_dual4} is equivalent to
\eqref{AppB_reduced_original4} and therefore gives no additional
condition. The remaining equations determine the compatibility
conditions imposed on $F$.

\medskip

\noindent First, subtracting \eqref{AppB_reduced_original1} from
\eqref{AppB_reduced_dual1} eliminates the common terms
$X_a(\alpha)$ and $i_{X_a}X_b(\alpha)$. Using
\begin{align}
e^b\wedge
(i_{X_a}i_{X_b}\omega)
&=
i_{X_a}\omega
-
i_{X_a}
\bigl(e^b\wedge i_{X_b}\omega\bigr),
\nonumber\\
e^b\wedge i_{X_b}\gamma
&=
q\,\gamma,
\qquad
\gamma\in\Omega^q,
\end{align}
and collecting the remaining terms gives precisely
\eqref{F0cond}. Thus the first compatibility condition is
$(F0)_a=0$.

\medskip

\noindent Next, we use \eqref{AppB_reduced_original3} and
\eqref{AppB_reduced_original4} in
\eqref{AppB_reduced_dual3} to eliminate
$e^b\wedge X_b(\beta)$ and $X_\theta(\beta)$. This gives
\begin{equation}\label{AppB_F1cond}
\begin{aligned}
0={}&
p\,d\sigma\wedge F
+
e^b\wedge X_b(F)
\\
&\quad
-
e^{-\sigma}
\Big[
2(p-1)e^{-\sigma}d\sigma\wedge\beta
-
\Big(
i_{X^m}(dA+db)
+
(i_{X^m}d\sigma)(A-b)
\Big)\wedge i_{X_m}\alpha
\Big].
\end{aligned}
\end{equation}
We denote the left-hand side of \eqref{AppB_F1cond} by $F1$.

\medskip

\noindent Finally, we use \eqref{AppB_reduced_original2} in
\eqref{AppB_reduced_dual2} to remove the terms
$X_a(\beta)$ and $i_{X_a}X_b(\beta)$. The remaining
$X_\theta(\beta)$ contribution is eliminated by
\eqref{AppB_reduced_original4}, together with its contraction by
$i_{X_a}$. After collecting terms, the resulting equation is exactly
\eqref{F2cond}; hence the second compatibility condition is
$(F2)_a=0$.

\medskip

\noindent The relation $F1=0$ is not independent. Indeed, taking the
exterior trace of \eqref{F2cond} gives
\begin{equation}\label{AppB_Ftrace}
e^a\wedge(F2)_a=F1.
\end{equation}
Consequently, the independent compatibility conditions are
\begin{align}
(F0)_a=0,
\qquad
(F2)_a=0,
\end{align}
as stated in Section~\ref{sec3}.

\end{document}